\documentclass[aps,pra,superscriptaddress,amsmath,amssymb]{revtex4-1}
\usepackage{graphicx}

\begin{document}

\title{Exchange of optical vortices using an electromagnetically induced
transparency based four-wave mixing setup}

\author{Hamid Reza Hamedi}
\email{hamid.hamedi@tfai.vu.lt}

\affiliation{Institute of Theoretical Physics and Astronomy, Vilnius University,
Saul\.etekio 3, Vilnius LT-10222, Lithuania}

\author{Julius Ruseckas}
\email{julius.ruseckas@tfai.vu.lt}

\affiliation{Institute of Theoretical Physics and Astronomy, Vilnius University,
Saul\.etekio 3, Vilnius LT-10222, Lithuania}

\author{Gediminas Juzeli\={u}nas}
\email{gediminas.juzeliunas@tfai.vu.lt}

\affiliation{Institute of Theoretical Physics and Astronomy, Vilnius University,
Saul\.etekio 3, Vilnius LT-10222, Lithuania}
\begin{abstract}
We propose a scheme to exchange optical vortices of slow light using
the phenomenon of electromagnetically induced transparency (EIT) in
a four-level double-$\Lambda$ atom-light coupling scheme illuminated
by a pair of probe fields as well as two control fields of larger
intensity. We study the light-matter interaction under the situation
where one control field carries an optical vortex, and another control
field has no vortex. We show that the orbital angular momentum (OAM)
of the vortex control beam can be transferred to a generated probe
field through a four-wave mixing (FWM) process and without switching
on and off of the control fields. Such a mechanism of OAM transfer
is much simpler than in a double-tripod scheme in which the exchange
of vortices is possible only when two control fields carry optical
vortices of opposite helicity. The losses appearing during such OAM
exchange are then calculated. It is found that the one-photon detuning
plays an important role in minimizing the losses. An approximate analytical expression is obtained for the optimal one-photon detuning for which the losses are minimum while the intensity of generated probe field is maximum.
  The influence of phase mismatch on the exchange of optical vortices is also investigated.
We show that in presence of phase mismatch the exchange of optical vortices can still be efficient.
\end{abstract}

\pacs{42.50.Gy; 42.50.Ct; 42.50.-p}
\maketitle

\section{Introduction}

With the recent progress in quantum optics and laser physics, a significant
attention has been drawn to novel approaches \cite{Kawazoe2003,Jiang2004}
which enable coherent control of interaction between radiation and
matter. A promising and flexible technique to manipulating pulse propagation
characteristics in atomic structures involves the quantum interference
and coherence \cite{Fleischhauer2005EIT,Fleischhauer1992coherence,Harris1990,Zibrov1995Interference,Harris1997,Fleischhauer2000}.
It turns out that the quantum interference between different excitation
channels is the basic mechanism for modifying optical response of
the medium to the applied fields, allowing to control the optical
properties of the medium. A destructive quantum interference in an
absorbing medium can suppress the absorption of the medium. This effect
has given the name Electromagnetically induced transparency (EIT)
\cite{Fleischhauer2000,Harris1997,Fleischhauer2005EIT}. An EIT medium
can be very dispersive resulting in a slowly propagating beam of the
electromagnetic radiation. The slow light \cite{Sahrai2004sublominal,Ruseckas2011,Ruseckas2007,Fleischhauer2016,Hamedi2017},
forming due to the EIT can greatly enhance the atom-light interaction
resulting in a number of distinctive optical phenomena \cite{Schmidt1996,Sahrai2005,Harshawardhan1996,imamoglu1997,LWIzhu,Hamedi2015,Hamedi2016}.

On the other hand, orbital angular momentum (OAM) of light provides
additional possibilities in manipulating the light propagation characteristics
\cite{AllenOAM1999,Alison2011OAM}. The OAM also represents an extra
degree of freedom in controlling the slow light \cite{Ruseckas2007,Ruseckas2011,Ruseckas2013transfer},
which can be exploited in quantum computation and quantum information
storage \cite{Allen2003book}.

Most scenarios considered earlier deals with the three-level $\Lambda$-type
level structure in which the incident probe field has a vortex \cite{Dutton2004,Pugatch2007,Wang2008OAM,Wang2008VORTEX,Moretti2009},
yet the control beam does not carry OAM. If the control beam carries
an OAM, the EIT is destroyed resulting in the absorption losses at
the vortex core. This is due to the zero intensity at the core of
the vortex beam. To avoid such losses, a four-level atom-light coupling
tripod-type configuration \cite{Unanyan1998,Gavra2007,PhysRevA.75.013810,Mazets2005,Ruseckas2005,Wang2004,Petrosyan2004,Rebi2004,Paspalakis2002}
was suggested with an extra control laser beam without an optical
vortex \cite{Ruseckas2011}. The total intensity of the control lasers
is then nonzero at the vortex core of the first control laser thus
avoiding the losses. It was shown that the OAM of the control field
can be transferred to the probe field in such a medium during switching
off and on the control beams \cite{Ruseckas2011}. Later a more complex
double tripod (DT) scheme of the atom-light coupling with six laser
fields was employed to transfer of the vortex between the control
and probe beams without switching off and on the control beams \cite{Ruseckas2013transfer}.
However, the transfer of the optical vortex in the DT scheme takes
place only when two control beams carry optical vortices of opposite
helicity. In this paper we consider a much simpler four-level double-$\Lambda$
(DL) EIT scheme for the exchange of optical vortices.

The DL atom\textendash light coupling scheme has attracted a great
deal of attention due to its important applications in coherent control
of pulse propagation characteristics \cite{DengPRE2005,Hoonsoo2011,Korsunsky1999DL,Zhuan2006,arXiv:1409.4153,arXiv:1302.1744v2,Liu2004,Huss2002DLphase,Liu2016DL,Wang2006DL,raczyski2004,Huang2006DL,Kang2006DL,Xiwei2013,wo2008magnts,Moiseev2006,Payne2002DL,Bongjune2013,Sahrai2009DL,Chong2008DL,Shpaisman2005DL,Chiu2014,LossMcCormick,LossTurnbullDL,arxiv.org/abs/1802.08874}.
All involved transitions in the DL scheme are excited by laser radiation
making a closed-loop phase sensitive coherent coupling scheme. It
was shown that an interference of excitation channels in such closed-loop
systems results in a strong dependence of the atomic state on the
relative phase and the relative amplitudes of the applied fields \cite{Korsunsky1999DL}.
As a result, the response of such a medium may be controlled by the
relative phase. Phase control of EIT in DL scheme has been investigated
in details both theoretically \cite{arXiv:1409.4153} and experimentally
\cite{Liu2016DL}. The EIT as well as light storing have been discussed
in the case of two signal pulses propagating in a four-level DL scheme
\cite{raczyski2004}. Observation of interference between three-photon
and one-photon excitations, and phase control of light attenuation
or transmission in a medium of four-level atoms in DL configuration
was reported by Kang et al. \cite{Kang2006DL}. Moiseev and Ham have
presented the two-color stationary light and quantum wavelength conversion
using double quantum coherence in a DL atomic structure \cite{Moiseev2006}.
All-optical image switching was demonstrated in a DL system, with
optical images generated by two independent laser sources \cite{Bongjune2013}.
It has been shown both analytically and numerically that a DL scheme
characterized by parametric amplification of cross-coupled probe and
four-wave mixing pulses, is an excellent medium for producing both
slow and stored light \cite{Eilam2008DL}. 

In this paper we employ DL level structure in order to exchange of
optical vortices between control and probe fields. It is shown that
the OAM of the control field can be transferred from the control field
to a generated probe field through a four-wave mixing (FWM) process
and without switching on and off of the control fields. We calculate
the losses appearing during such vortex exchange and obtain an approximate
analytical expression for the optimal one-photon detuning for which
the intensity of generated probe beam is maximum and the losses are
minimum. 

\section{Theoretical model and formulation\label{sec:Formulation-and-theoretical}}

\subsection{The double-$\Lambda$ system}

\begin{figure}
\includegraphics[width=0.5\columnwidth]{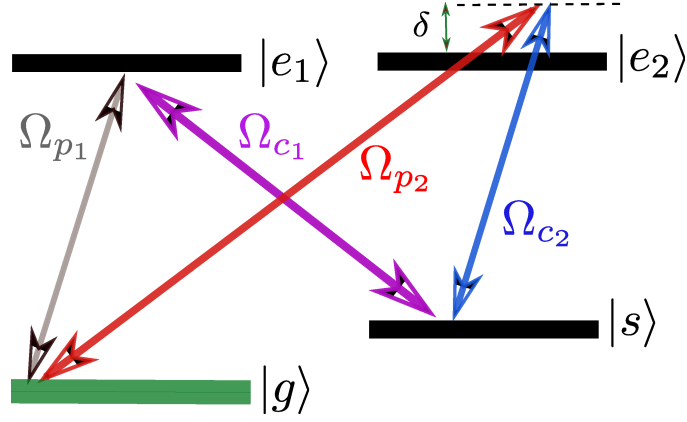} \caption{Schematic diagram of the four-level Double-$\Lambda$ (DL) quantum
system.}
\label{fig:1}
\end{figure}

We shall analyze the light-matter interaction in an ensemble of atoms using a
four-level double-$\Lambda$ (DL) scheme shown in Fig.~\ref{fig:1}.  The atoms
are characterized by two metastable ground states $|g\rangle$ and $|s\rangle$,
as well as two excited states $|e_{1}\rangle$ and $|e_{2}\rangle$. The scheme
is based on a mixture of two EIT subsystems in $\Lambda$ configuration. The
first $\Lambda$ subsystem is formed by a weak probe field described by a Rabi
frequency $\Omega_{p_{1}}$ and a strong control field with a Rabi-frequency
$\Omega_{c_{1}}$.  Another weak probe field with a Rabi frequency
$\Omega_{p_{2}}$ and a strong field with the Rabi-Frequency $\Omega_{c_{2}}$
build the second $\Lambda$ subsystem. For each EIT subsystem, the strong laser
fields $\Omega_{c_{1}}$ and $\Omega_{c_{2}}$ control the propagation of probe
fields $\Omega_{p_{1}}$and $\Omega_{p_{2}}$ through the medium inducing a
transparency for the resonant probe beams due to the destructive quantum
interference \cite{Moiseev2006}. We consider the situation when all the light
beams are co-propagating in the same direction.  All involved transitions are
then excited by laser radiation making a closed-loop phase sensitive coherent
coupling scheme described by the Hamiltonian ($\hbar=1$)
\begin{equation}
H_{\mathrm{DL}}=-\Omega_{p_{1}}|e_{1}\rangle\langle g|-\Omega_{p_{2}}|e_{2}\rangle\langle g|-\Omega_{c_{1}}|e_{1}\rangle\langle s|-\Omega_{c_{2}}|e_{2}\rangle\langle s|+\mathrm{H.c.}\,.\label{eq:1}
\end{equation}
Obviously, the whole DL atom-light coupling scheme then includes two
FWM pathways $|g\rangle\rightarrow|e_{2}\rangle\rightarrow|s\rangle\rightarrow|e_{1}\rangle\rightarrow|g\rangle$
generating the first probe beam and $|g\rangle\rightarrow|e_{1}\rangle\rightarrow|s\rangle\rightarrow|e_{2}\rangle\rightarrow|g\rangle$
creating the second probe beam.

\subsection{Maxwell-Bloch equations}

Now we begin to derive the basic equations describing the interaction
between optical fields and DL atoms. We shall neglect the atomic center-of-mass
motion. We assume that both probe fields $\Omega_{p_{1}}$and $\Omega_{p_{2}}$
are much weaker than the control fields $\Omega_{c_{1}}$ and $\Omega_{c_{2}}$.
As a result, all the atoms remain in the ground state $|g\rangle$
and one can treat the contribution of the probe fields as a perturbation
in the derivation of the following equations:
\begin{align}
\dot{\rho}_{e_{1}g}= & -\frac{\gamma_{e_{1}}}{2}\rho_{e_{1}g}+\frac{i}{2}\Omega_{c_{1}}\rho_{sg}+\frac{i}{2}\Omega_{p_{1}}\,,\label{eq:2}\\
\dot{\rho}_{e_{2}g}= & \left(i\delta-\frac{\gamma_{e_{2}}}{2}\right)\rho_{e_{2}g}+\frac{i}{2}\Omega_{c_{2}}\rho_{sg}+\frac{i}{2}\Omega_{p_{2}}\,,\label{eq:3}\\
\dot{\rho}_{sg}= & \frac{i}{2}\Omega_{c_{1}}^{*}\rho_{e_{1}g}+\frac{i}{2}\Omega_{c_{2}}^{*}\rho_{e_{2}g}\,,\label{eq:4}
\end{align}  
and
\begin{align}
\frac{\partial\Omega_{p_{1}}}{\partial z}+c^{-1}\frac{\partial\Omega_{p_{1}}}{\partial t} & =i\frac{\alpha_{p_{1}}\gamma_{e_{1}}}{2L}\rho_{e_{1}g}\,,\label{eq:5}\\
\frac{\partial\Omega_{p_{2}}}{\partial z}+c^{-1}\frac{\partial\Omega_{p_{2}}}{\partial t} & =i\frac{\alpha_{p_{2}}\gamma_{e_{2}}}{2L}\rho_{e_{2}g}\,,\label{eq:6}
\end{align}
where $\rho_{u,v}$ are the matrix elements of the density matrix
operator $\rho=\sum_{u,v}|u\rangle\rho_{uv}\langle v|$, $\gamma_{e_{1}}$
and $\gamma_{e_{2}}$ represent the total decay rates from the excited
states $|e_{1}\rangle$ and $|e_{2}\rangle$, $\alpha_{p_{1}}$ and
$\alpha_{p_{2}}$ denote the optical depth of first and second probe
fields, $L$ describes the optical length of the medium, and $\delta=\omega_{d}-\omega_{24}$
is the detuning of the driving transition, where $\omega_{d}$ and
$\omega_{se_{2}}$ are the frequencies of driving field and the $|s\rangle\leftrightarrow|e_{2}\rangle$
transition. It should be noted that the diffraction terms containing
the transverse derivatives $\nabla_{\perp}^{2}\Omega_{p_{1}}$and
$\nabla_{\perp}^{2}\Omega_{p_{2}}$ have been neglected in the Maxwell
equations (\ref{eq:5}) and (\ref{eq:6}). These terms are negligible
if the phase change of the probe fields due to these terms are much
smaller than $\pi$. One can estimate them as $\nabla_{\perp}^{2}\Omega_{p_{1(2)}}\sim w^{-2}\Omega_{p_{1(2)}}$,
where $w$ indicates a characteristic transverse dimension of the
probe beams (representing a width of the vortex core if the probe
beam carries an optical vortex, or a characteristic width of the beam
if the probe beam has no optical vortex). The temporal change of the
probe field can be approximated as $\frac{\partial\Omega_{p_{1(2)}}}{\partial t}\sim cL^{-1}\Delta\Omega_{p_{1(2)}}$,
where $\Delta\Omega_{p_{1(2)}}$ is the change of the field and $L$
shows the length of the atomic sample. Therefore, the change of the
phase due to the diffraction term could be $L/2kw^{2}$, where $k=\mathrm{diag}(k_{1},k_{2})$
is a diagonal $2\times2$ matrix of the wave vectors of the probe
field, with $k_{i}=\omega_{i}/c$ being the central wave vector of
the $i$th probe beam. The latter can be neglected when the sample
length $L$ is not too large, $L\lambda/w^{2}\ll1$. Taking the length
of the atomic cloud $L=100\,\mathrm{\mu m}$, the characteristic transverse
dimension of the probe beam $w=20\,\mathrm{\mu m}$ and the wavelength
$\lambda=1\,\mathrm{\mu m}$, we obtain $L\lambda/w^{2}=0.25$. Therefore
we can drop out the diffraction term, yielding Eqs.~(\ref{eq:5})
and (\ref{eq:6}).

\section{Analytical solutions for two probe beams propagation}

We consider the situation when envelopes of all interacting pulses are long
and flat. Thus, except for a short transient period, the envelopes can be
assumed to be constant most of the time.  Hence one can consider the steady
state of the fields by dropping the time derivatives in the equations. To
simplify the discussion, let us take $\alpha_{p_{1}}=\alpha_{p_{2}}=\alpha$ and
$\gamma_{e_{1}}=\gamma_{e_{2}}=\Gamma$. Using Eqs.~(\ref{eq:2})--(\ref{eq:4}),
one then arrives at the simple analytical expressions for the steady-state
solutions of $\rho_{e_{1}g}$ and $\rho_{e_{2}g}$
\begin{align}
\rho_{e_{1}g}= & \frac{\Omega_{p_{2}}\Omega_{c_{1}}\Omega_{c_{2}}^{*}-\Omega_{p_{1}}|\Omega_{c_{2}}|^{2}}{2\delta|\Omega_{c_{1}}|^{2}+i\Gamma|\Omega|^{2}}\,,\label{eq:7}\\
\rho_{e_{2}g}= & \frac{\Omega_{p_{1}}\Omega_{c_{2}}\Omega_{c_{1}}^{*}-\Omega_{p_{2}}|\Omega_{c_{1}}|^{2}}{2\delta|\Omega_{c_{1}}|^{2}+i\Gamma|\Omega|^{2}}\,,\label{eq:8}
\end{align}
where 
\begin{equation}
|\Omega|^{2}=|\Omega_{c_{1}}|^{2}+|\Omega_{c_{2}}|^{2}
\end{equation}
is the total Rabi-frequency of the control fields. If the second probe
field is zero $\Omega_{p_{2}}(0)=0$ at the entrance ($z=0$), substituting
Eqs.~(\ref{eq:7}) and (\ref{eq:8}) into the Maxwell equations (\ref{eq:5})
and (\ref{eq:6}) results in
\begin{align}
\Omega_{p_{1}}(z)= & \frac{\Omega_{p_{1}}(0)}{|\Omega|^{2}}\left(|\Omega_{c_{1}}|^{2}+|\Omega_{c_{2}}|^{2}\exp\left(-i\frac{\alpha z}{2Ld}\right)\right)\,,\label{eq:9}\\
\Omega_{p_{2}}(z)= & \frac{\Omega_{p_{1}}(0)}{|\Omega|^{2}}\Omega_{c_{1}}^{*}\Omega_{c_{2}}\left(1-\exp\left(-i\frac{\alpha z}{2Ld}\right)\right)\,,\label{eq:10}
\end{align}
with $d=i+2|\Omega_{c_{1}}|^{2}\delta/(\Gamma|\Omega|^{2})$, where
$\Omega_{p_{1}}(0)$ represents the incident first probe beam. 

\section{Exchange of optical vortices}

\begin{figure}
\includegraphics[width=0.5\columnwidth]{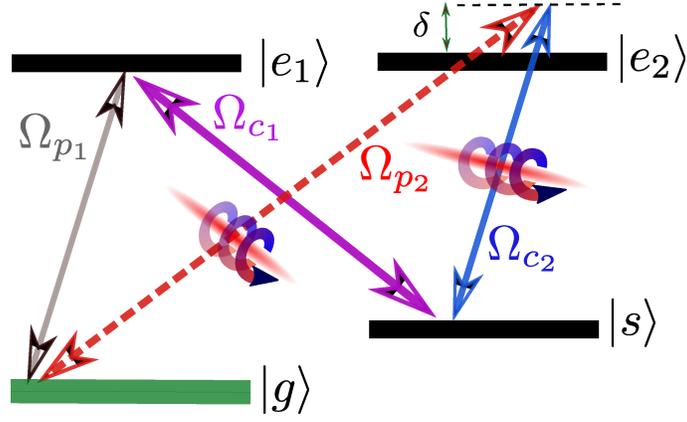} 

\caption{Transfer of OAM from the control field $\Omega_{c_{2}}$ to the second
generated probe field $\Omega_{p_{2}}$.}
\label{fig:2}
\end{figure}

Let us allow the one of the control field photons to have an orbital
angular momentum $\hbar l$ along the propagation axis $z$ \cite{AllenOAM1999}.
In that case the second control field $\Omega_{c_{2}}$ is characterized
by the Rabi-frequency
\begin{equation}
\Omega_{c_{2}}=|\Omega_{c_{2}}|\exp(il\Phi),\label{eq:15}
\end{equation}
where $\Phi$ is the azimuthal angle. For a Laguerre-Gaussian (LG)
doughnut beam we may write 
\begin{equation}
|\Omega_{c_{2}}|=\varepsilon_{c_{2}}\left(\frac{r}{w}\right)^{|l|}\exp\left(-\frac{r^{2}}{w^{2}}\right)\,,\label{eq:16}
\end{equation}
where $r$ represents the distance from the vortex core (cylindrical
radius), $w$ denotes the beam waist parameter, and $\varepsilon_{c_{2}}$
is the strength of the vortex beam. The Rabi-frequency of the first
control field does not have a vortex and is given by
\begin{equation}
\Omega_{c_{1}}=|\Omega_{c_{1}}|\,.\label{eq:17}
\end{equation}
Substituting Eqs.~(\ref{eq:15})-(\ref{eq:17}) into Eqs.~(\ref{eq:9})
and (\ref{eq:10}), we obtain
\begin{align}
\Omega_{p_{1}}(z)= & \Omega_{p_{1}}(0)\frac{\left(|\Omega_{c_{1}}|^{2}+\varepsilon_{c_{2}}^{2}\left(\frac{r}{w}\right)^{2|l|}\exp\left(-\left(2\frac{r^{2}}{w^{2}}+i\frac{\alpha z}{2Ld}\right)\right)\right)}{|\Omega_{c_{1}}|^{2}+\varepsilon_{c_{2}}^{2}\left(\frac{r}{w}\right)^{2|l|}\exp\left(-2\frac{r^{2}}{w^{2}}\right)}\,,\label{eq:18}\\
\Omega_{p_{2}}(z)= & \Omega_{p_{1}}(0)\left(\frac{r}{w}\right)^{|l|}\exp(il\Phi)\frac{|\Omega_{c_{1}}|\varepsilon_{c_{2}}\exp\left(-\frac{r^{2}}{w^{2}}\right)\left(1-\exp\left(-i\frac{\alpha z}{2Ld}\right)\right)}{|\Omega_{c_{1}}|^{2}+\varepsilon_{c_{2}}^{2}\left(\frac{r}{w}\right)^{2|l|}\exp\left(-2\frac{r^{2}}{w^{2}}\right)}\,.\label{eq:19}
\end{align}
In this way, the second probe field $\Omega_{p_{2}}\sim\exp(il\Phi)$
is generated with the the same vorticity as the control field $\Omega_{c_{2}}\sim\exp(il\Phi)$,
as illustrated in Fig.~\ref{fig:2}. Also, an extra factor $\propto r^{|l|}$
in Eq.~(\ref{eq:19}) indicates that in the vicinity of the vortex
core the generated probe beam looks like a LG doughnut beam, with
the intensity going to zero for $r\rightarrow0$. Note that due to
the presence of a non-vortex control beam $\Omega_{c_{1}}$, the total
intensity of the control lasers is not zero at the vortex core, preventing
the absorption losses.

\begin{figure}
\includegraphics[width=0.35\columnwidth]{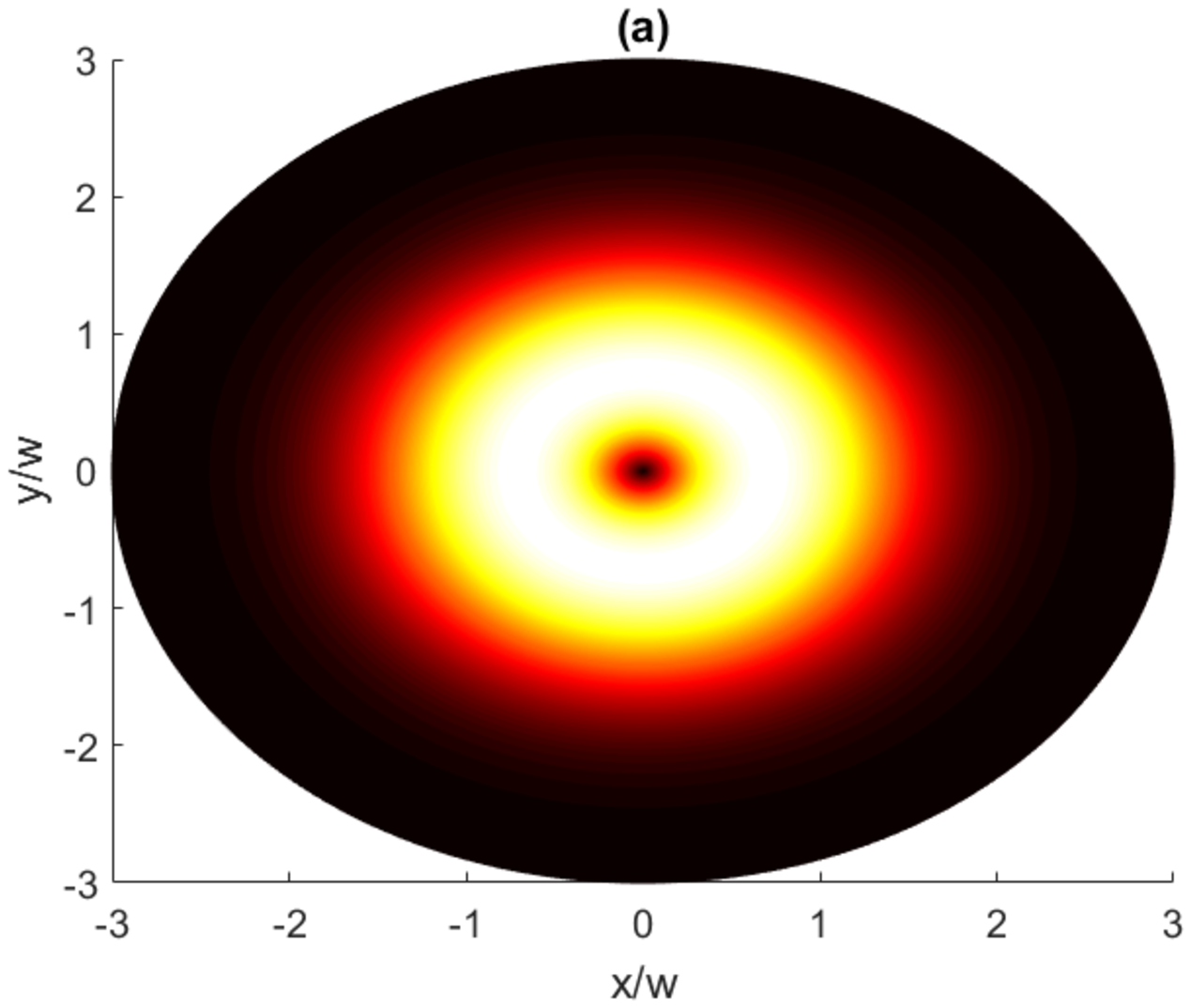} ~~\includegraphics[width=0.35\columnwidth]{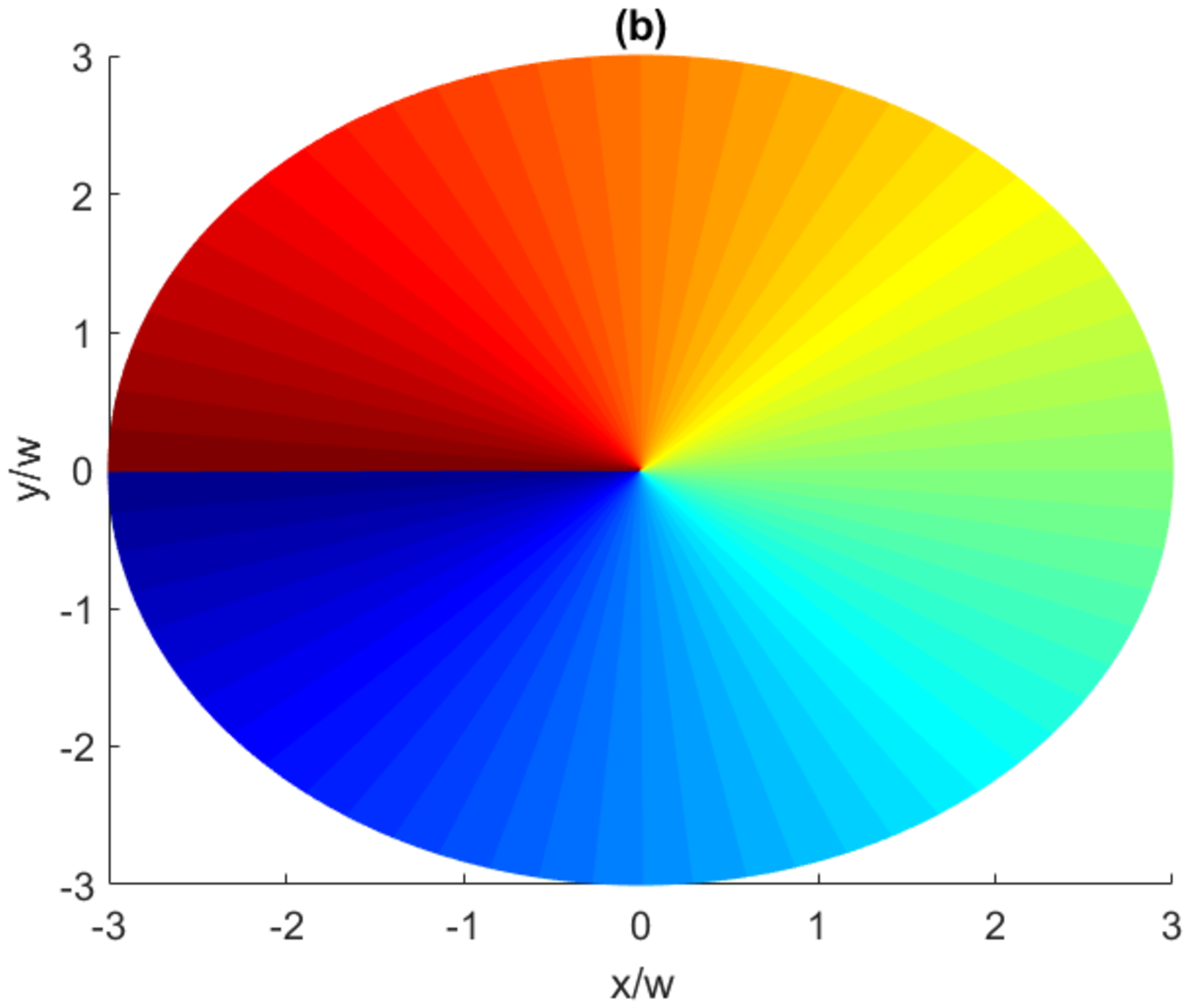} 

\includegraphics[width=0.35\columnwidth]{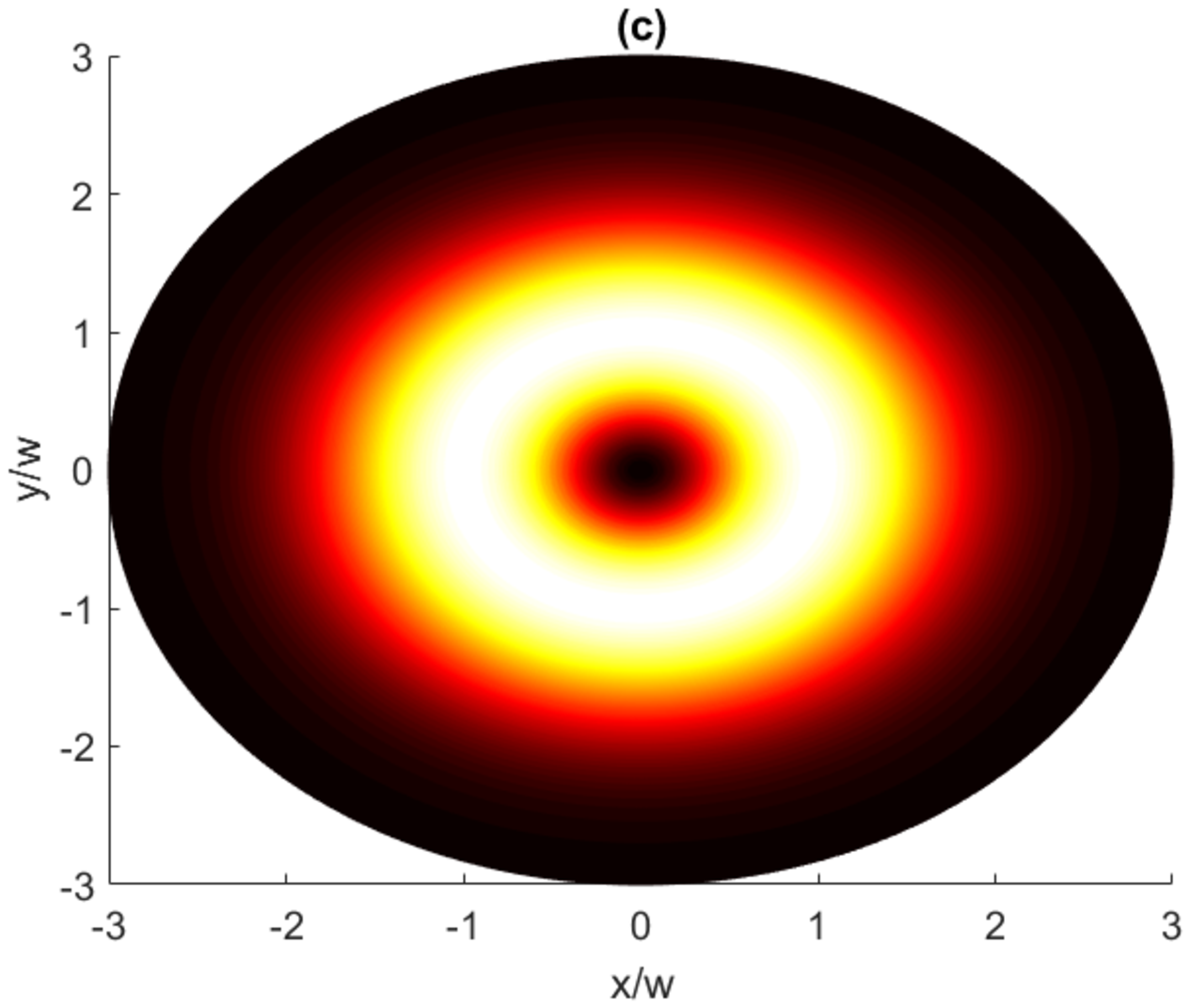} ~~\includegraphics[width=0.35\columnwidth]{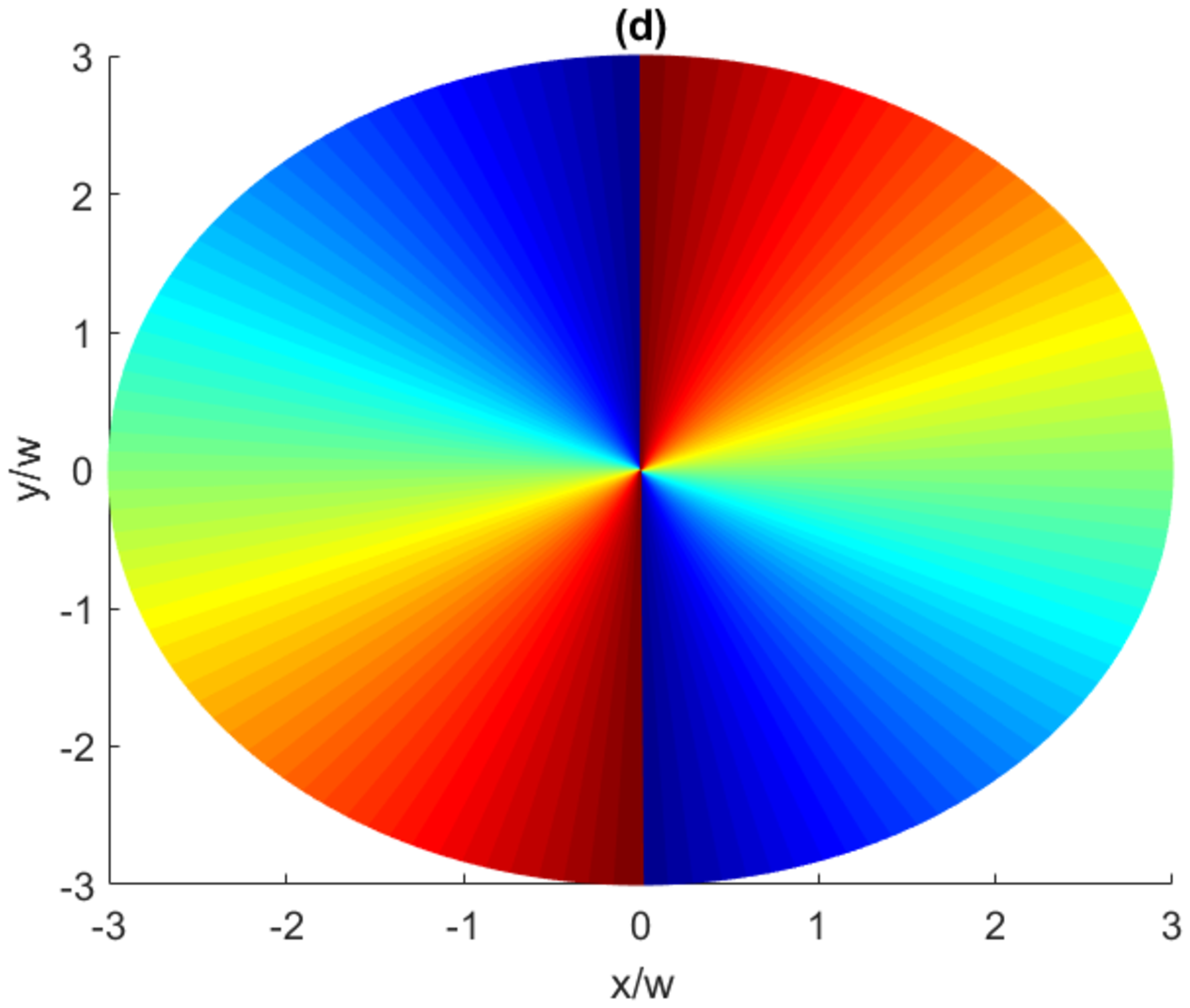} 

\includegraphics[width=0.35\columnwidth]{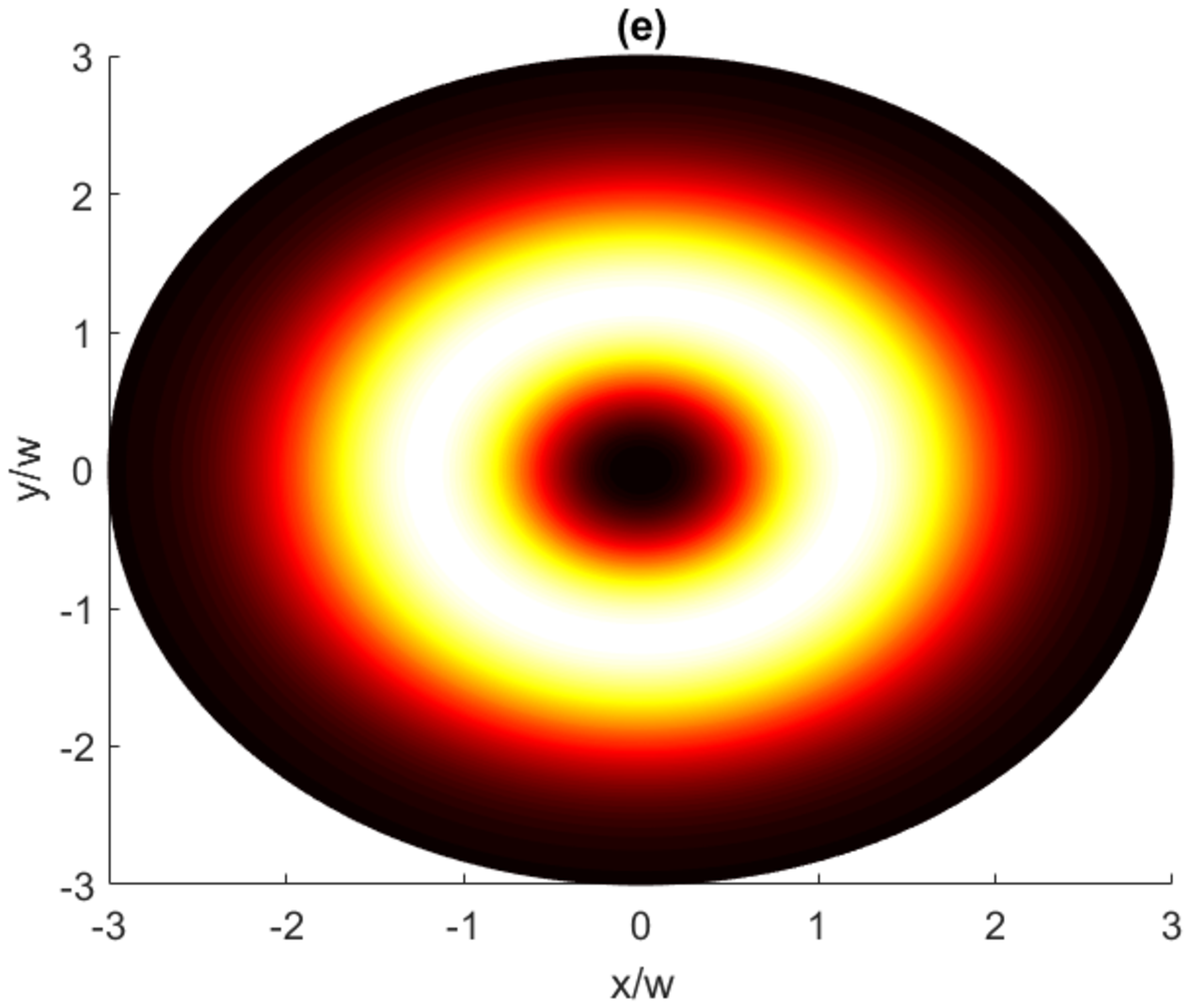} ~~\includegraphics[width=0.35\columnwidth]{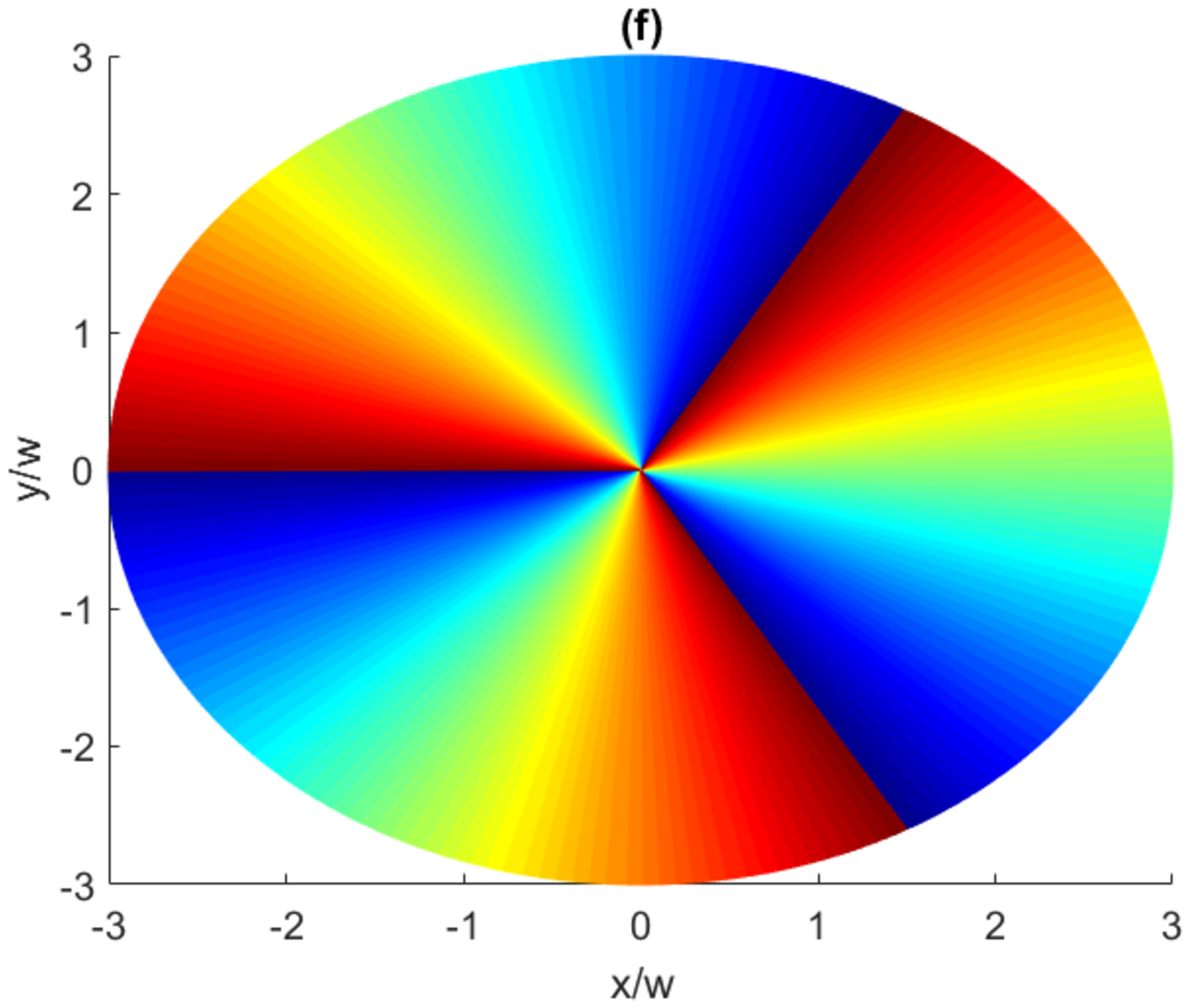} 

\caption{Intensity distributions (a, c, e) in arbitrary units as well as the corresponding helical
phase patterns (b, d, f) of the generated second probe vortex beam
$\Omega_{p_{2}}$ with different OAM numbers $l=1$ (a, b), $l=2$
(c, d) and $l=3$ (e, f). Here the parameters are $|\Omega_{c_{1}}|=\varepsilon_{c_{2}}=\Gamma$,
$\delta=0$, $\Omega_{p_{1}}(0)=0.1\Gamma$, $z=L$ and $\alpha=100$. The position
is plotted in dimensionless units.}
\label{fig:3}
\end{figure}

The intensity distributions and the corresponding helical phase pattern
of the generated second probe vortex beam $\Omega_{p_{2}}$ are shown
in Fig.~\ref{fig:3} for different topological charge numbers and
under the resonance condition $\delta=0$. When $l=1$, a doughnut
intensity profile is observed with a dark hollow center (Fig.~\ref{fig:3}
(a)). The phase pattern corresponding to this case is plotted in Fig.~\ref{fig:3}(b).
One can see that the phase jumps from $0$ to $2\pi$ around the singularity
point. When the OAM number $l$ increases to larger number $l=2$
($l=3$), the dark hollow center is increased in size as shown in
Fig.~\ref{fig:3}(c) (Fig.~\ref{fig:3}(e)), while the phase jumps
from $0$ to $4\pi$ ($6\pi$) around the singularity point, as can
be seen in Fig.~\ref{fig:3}(d) (Fig.~\ref{fig:3}(f)). 

\section{Estimation of losses}

At the beginning of the atomic medium where the first probe beam $\Omega_{p_{1}}$
has just entered, the second probe beam has not yet been generated
($\Omega_{p_{2}}(0)=0$), so it is not yet driving the transition
$|g\rangle\leftrightarrow|e_{2}\rangle$. In this situation, the four-level
DL level scheme reduces to an $N$-type atom-light coupling scheme
for which a strong absorption is expected \cite{Jiteng2011N}. Going
deeper into the atomic medium, the second probe beam is created (see
Eq.~(\ref{eq:19})), resulting in reduction of absorption losses.
Yet the losses by absorption are nonzero. It has been shown that the
non-degenerate forward FWM in the DL medium can generate strong quantum
correlations between twin beams of light in the DL level structure
which minimizes the absorption losses \cite{LossMcCormick}. The amount
of phase mismatch in the DL system has been also demonstrated to play
an important role to achieve the physical situations in which these
losses are minimal \cite{LossTurnbullDL}. In what fallows we will
estimate such energy losses, aiming at reaching an optimum condition
when the efficiency of the generated probe field is maximum whereas
the losses are minimum. Substituting Eqs.~(\ref{eq:18}) and (\ref{eq:19})
into Eqs.~(\ref{eq:7}) and (\ref{eq:8}), one gets 
\begin{align}
\rho_{e_{1}g}=&\frac{-\Omega_{p_{1}}(0)}{i\Gamma\left(|\Omega_{c_{1}}|^{2}+\varepsilon_{c_{2}}^{2}\left(\frac{r}{w}\right)^{2|l|}\exp\left(-2\frac{r^{2}}{w^{2}}\right)\right)+2\delta|\Omega_{c_{1}}|^{2}}\varepsilon_{c_{2}}^{2}\left(\frac{r}{w}\right)^{2|l|}\exp\left(-\left(2\frac{r^{2}}{w^{2}}+i\frac{\alpha z}{2Ld}\right)\right),\label{eq:22}\\
\rho_{e_{2}g}=&\frac{\Omega_{p_{1}}(0)}{i\Gamma\left(|\Omega_{c_{1}}|^{2}+\varepsilon_{c_{2}}^{2}\left(\frac{r}{w}\right)^{2|l|}\exp\left(-2\frac{r^{2}}{w^{2}}\right)\right)+2\delta|\Omega_{c_{1}}|^{2}}|\Omega_{c_{1}}|\varepsilon_{c_{2}}\left(\frac{r}{w}\right)^{|l|}\exp\left(-\left(\frac{r^{2}}{w^{2}}+i\frac{\alpha z}{2Ld}\right)\right)\exp(il\Phi).\label{eq:23}
\end{align}
The density matrix elements $\rho_{e_{1}g}$ and $\rho_{e_{2}g}$
generally describe the amplitudes of the excited states $|e_{1}\rangle$
and $|e_{2}\rangle$, respectively. Therefore Eqs.~(\ref{eq:22})
and (\ref{eq:23}) represent the occupation of excited states inside
the medium. It is seen that the losses appear mostly at the beginning
of the medium, when going through the medium the losses reduce. In
addition, when $r\rightarrow0$ both Eqs.~(\ref{eq:22}) and (\ref{eq:23})
are zero indicating that at the vortex core we have no losses. However,
away from the vortex core, these equations indicate that the excited
states $|e_{1}\rangle$ and $|e_{2}\rangle$ are occupied resulting
in absorption losses. Also it follows from Eqs.~(\ref{eq:22}) and
(\ref{eq:23}) that the one-photon detuning $\delta$ plays a critical
role in reducing the losses. For the detuning much larger than $\Gamma$
the denominator of Eqs.~(\ref{eq:22}) and (\ref{eq:23}) becomes
very large making negligible occupation of excited states and giving
rise to very small losses. However, a large detuning alone may make
creation of second probe field less efficient, as indicated by Eq.~(\ref{eq:19}).
The latter equation demonstrates that the FWM efficiency can be enhanced
by increasing the optical density $\alpha$ . One can conclude that
the large detuning requires larger optical density $\alpha$. Yet
the optical density $\alpha$ of the atomic cloud is constrained due
to neglecting the diffraction terms in the wave equations (\ref{eq:5})
and (\ref{eq:6}) \cite{Ruseckas2013transfer}. Since the optical
density $\alpha$ is proportional to the medium length $L$ and the
number density of atoms $n$, the restrictions on the length of the
medium limits also the optical density. 

It appears that for a given optical density $\alpha$ there is an
optimal value of one-photon detuning $\delta$ for which the intensity
of second probe field $|\Omega_{p_{2}}|^{2}$ is the maximum. In the
following we calculate an analytical expression for optimal $\delta$.
From Eq.~(\ref{eq:10}), it is straightforward to obtain the expression
for the real-valued quantity $|\Omega_{p_{2}}|^{2}$
\begin{equation}
|\Omega_{p_{2}}|^{2}=|\beta|^{2}\left[1+\exp\left(\frac{-2x}{1+y^{2}\delta^{2}}\right)-2\cos\frac{y\delta x}{1+y^{2}\delta^{2}}\exp\left(\frac{-x}{1+y^{2}\delta^{2}}\right)\right]\,,\label{eq:24}
\end{equation}
where $|\beta|^{2}=\frac{|\Omega_{p_{1}}(0)|^{2}}{|\Omega|^{4}}|\Omega_{c_{1}}|^{2}|\Omega_{c_{2}}|^{2}$,
$x=\alpha z/2L$, and $y=2|\Omega_{c_{1}}|^{2}/\Gamma|\Omega|^{2}$.
Calculating the derivative of Eq.~(\ref{eq:24}) with respect to
$\delta$ we get
\begin{equation}
\frac{\partial|\Omega_{p_{2}}|^{2}}{\partial\delta}=\frac{2xy|\beta|^{2}\exp\left(\frac{-2x}{1+y^{2}\delta^{2}}\right)\left(2y\delta+\exp\left(\frac{x}{1+y^{2}\delta^{2}}\right)\left(-2y\delta\cos\left(\frac{xy\delta}{1+y^{2}\delta^{2}}\right)+(1-y^{2}\delta^{2})\sin\left(\frac{xy\delta}{1+y^{2}\delta^{2}}\right)\right)\right)}{(1+y^{2}\delta^{2})^{2}}.\label{eq:25}
\end{equation}
The optimal detuning $\delta$ is found when
\begin{equation}
2y\delta+\exp\left(\frac{x}{1+y^{2}\delta^{2}}\right)\left(-2y\delta\cos\left(\frac{xy\delta}{1+y^{2}\delta^{2}}\right)+(1-y^{2}\delta^{2})\sin\left(\frac{xy\delta}{1+y^{2}\delta^{2}}\right)\right)=0\,.\label{eq:26}
\end{equation}
Assuming that $\delta$ is sufficiently large ($\delta\gg\Gamma$), we can
expand Eq.~(\ref{eq:26}) into Taylor series, yielding
\begin{equation}
\delta\approx\pm\frac{1}{\sqrt{6}}\frac{x}{y},\label{eq:27}
\end{equation}
or, in terms of the physical quantities,
\begin{equation}
\delta\approx\pm\frac{z\Gamma|\Omega|^{2}}{4\sqrt{6}L|\Omega_{c_{1}}|^{2}}\alpha.\label{eq:28}
\end{equation}

\begin{figure}
\includegraphics[width=0.5\columnwidth]{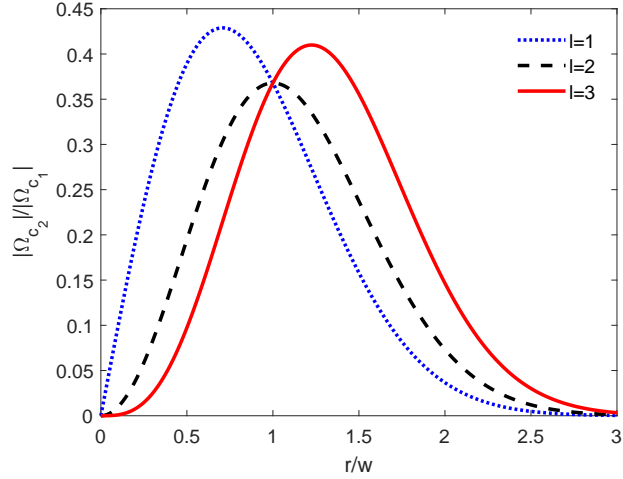} 
\caption{Dependence of the dimensionless quantity $|\Omega_{c_{2}}|/|\Omega_{c_{1}}|$ given in Eq.~(\ref{eq:16}) on
the dimensionless distance from the vortex core $r/w$ when $|\Omega_{c_{1}}|=\varepsilon_{c_{2}}=\Gamma$
and for different $l$ numbers.}
\label{fig:4}
\end{figure}

Equation (\ref{eq:28}) provides the values for optimal $\delta$
for which the efficiency of generation of $\Omega_{p_{2}}$ is the
largest. According to Eq.~(\ref{eq:28}) the optimal $\delta$ increases
linearly with the optical density $\alpha$. Since $\Omega_{c_{2}}$
represents an optical vortex, the ratio $|\Omega|^{2}/|\Omega_{c_{1}}|^{2}$
is not a constant and depends on $r$. In order to estimate an approximate
value for the optimal detuning $\delta$ we take the value of $|\Omega|^{2}/|\Omega_{c_{1}}|^{2}$
at the position $r$ of the maximum of $|\Omega_{c_{2}}|$. The dependence
of $|\Omega_{c_{2}}|/|\Omega_{c_{1}}|$  on the dimensionless
distance from the vortex core $r/w$ is plotted in Fig.~\ref{fig:4}
 by using Eq.~(\ref{eq:16}) and for different OAM numbers $l=1,2,3$. Subsequently the maximum of the quantity
$|\Omega_{c_{2}}|/|\Omega_{c_{1}}|$ ($|\Omega_{c_{2}}|_{max}/|\Omega_{c_{1}}|$) and the calculated
$|\Omega|^{2}/|\Omega_{c_{1}}|^{2}$ and $\delta$ at $|\Omega_{c_{2}}|_{\mathrm{max}}$
for different $l$ numbers are given in Table~\ref{tab:Table1}.
One can see, for example, that the maximum of $|\Omega_{c_{2}}|/|\Omega_{c_{1}}|$
for $l=1$ is about $0.42$. In this case, $|\Omega|^{2}/|\Omega_{c_{1}}|^{2}\approx1.17$,
yielding $\delta\approx\pm11.9\Gamma$. This is an approximate value
for the one-photon detuning $\delta$ for which the intensity of second
probe field $|\Omega_{p_{2}}|^{2}$ carrying an optical vortex with
$l=1$ is the largest, yet the losses are minimum. 

\begin{table}
\caption{The calculated values for $|\Omega_{c_{2}}|_{max}/|\Omega_{c_{1}}|$, $|\Omega|^{2}/|\Omega_{c_{1}}|^{2}$
and Optimal $\delta$ for different OAM numbers.  }

\begin{centering}
\begin{tabular}{cccc}
\hline 
 & $l=1$ & $l=2$ & $l=3$\tabularnewline
\hline 
$|\Omega_{c_{2}}|_{max}/|\Omega_{c_{1}}|$ & $0.42$ & $0.36$ & $0.4$\tabularnewline
$|\Omega|^{2}/|\Omega_{c_{1}}|^{2}$ & $1.17$ & $1.13$ & $1.16$\tabularnewline
Optimal $\delta$ & $\pm11.9\Gamma$ & $\pm11.53\Gamma$ & $\pm11.83\Gamma$\tabularnewline
\hline 
\end{tabular}
\par\end{centering}
\label{tab:Table1}
\end{table}

\begin{figure}
\includegraphics[width=0.35\columnwidth]{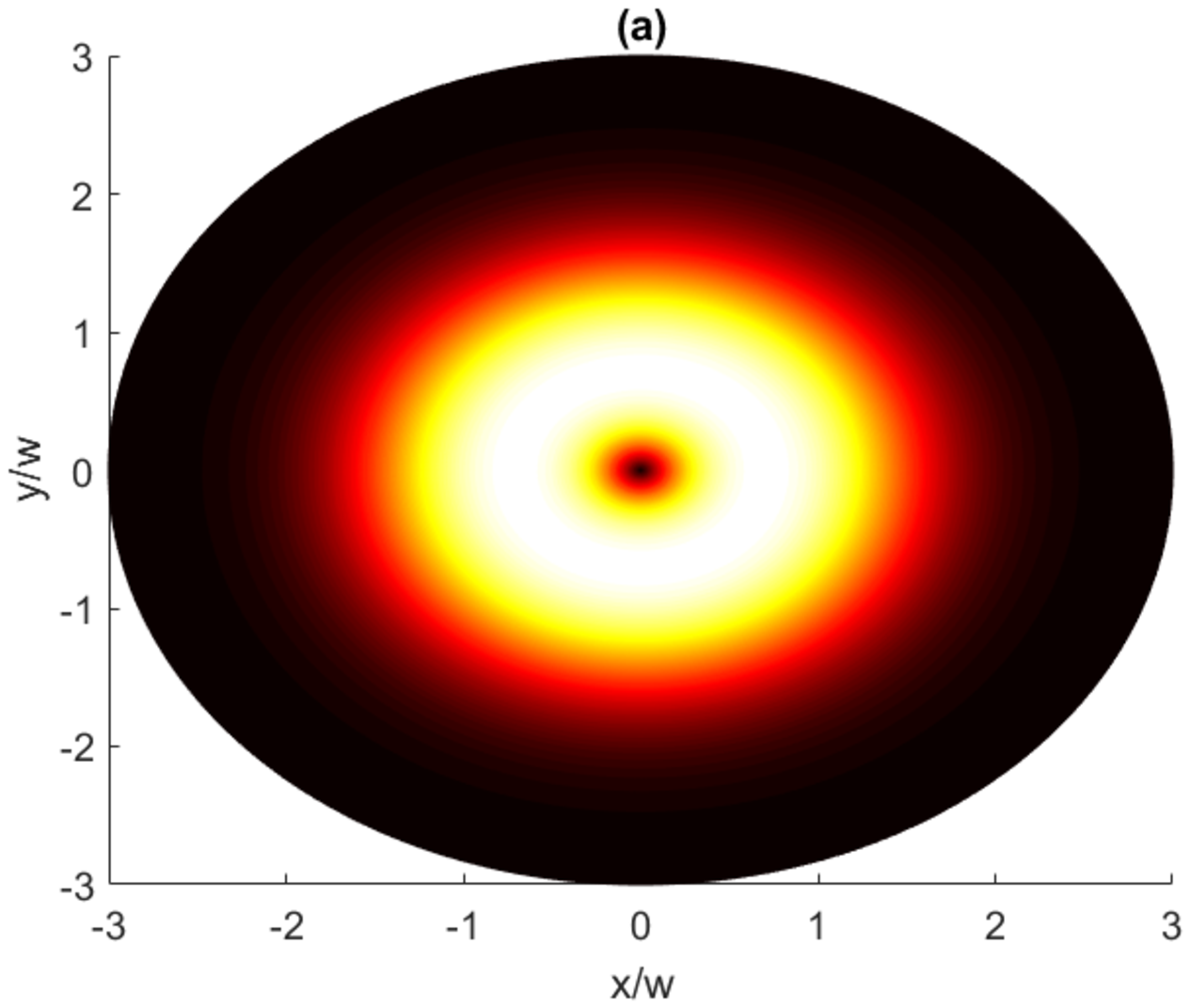} ~~\includegraphics[width=0.35\columnwidth]{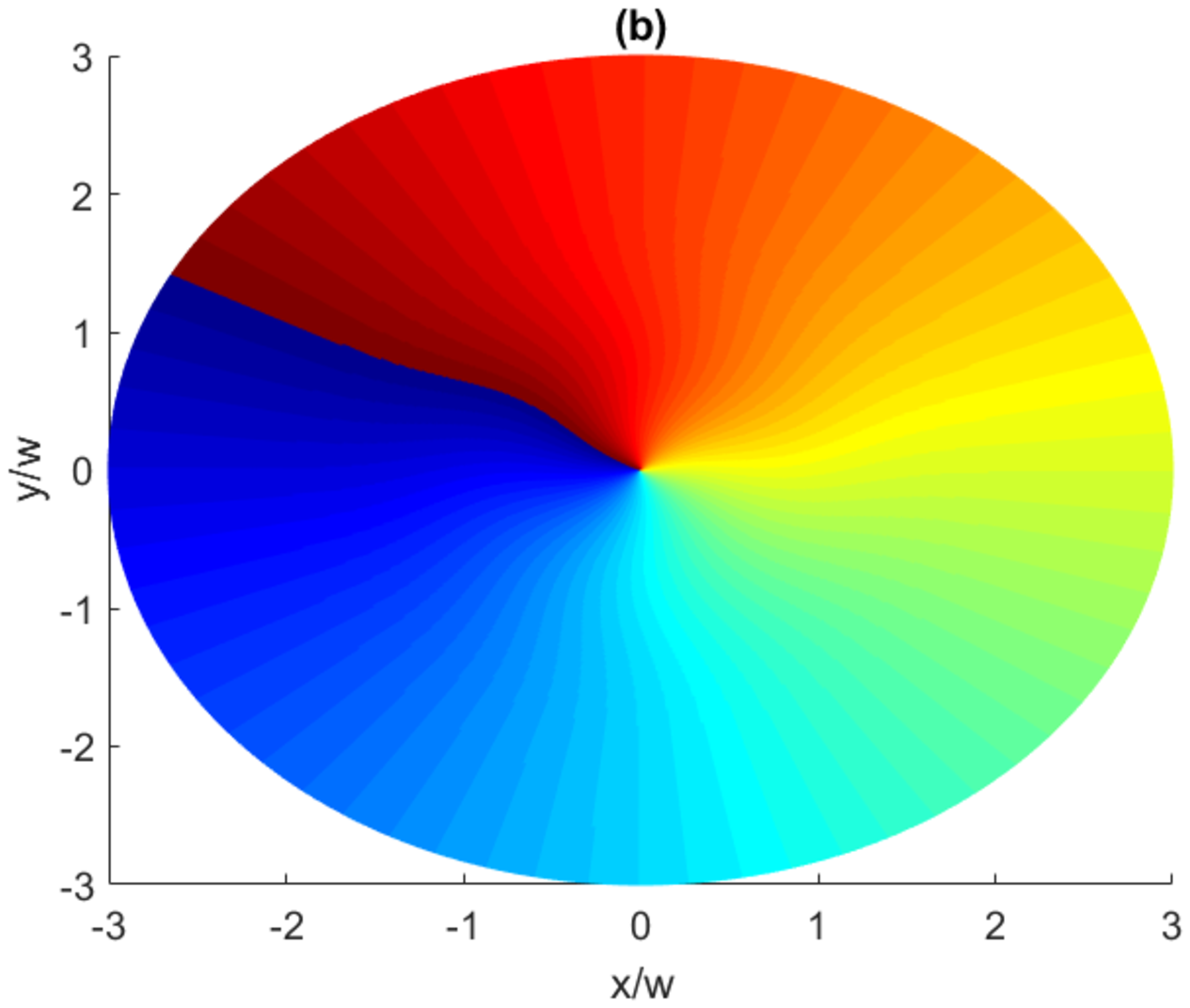} 

\includegraphics[width=0.35\columnwidth]{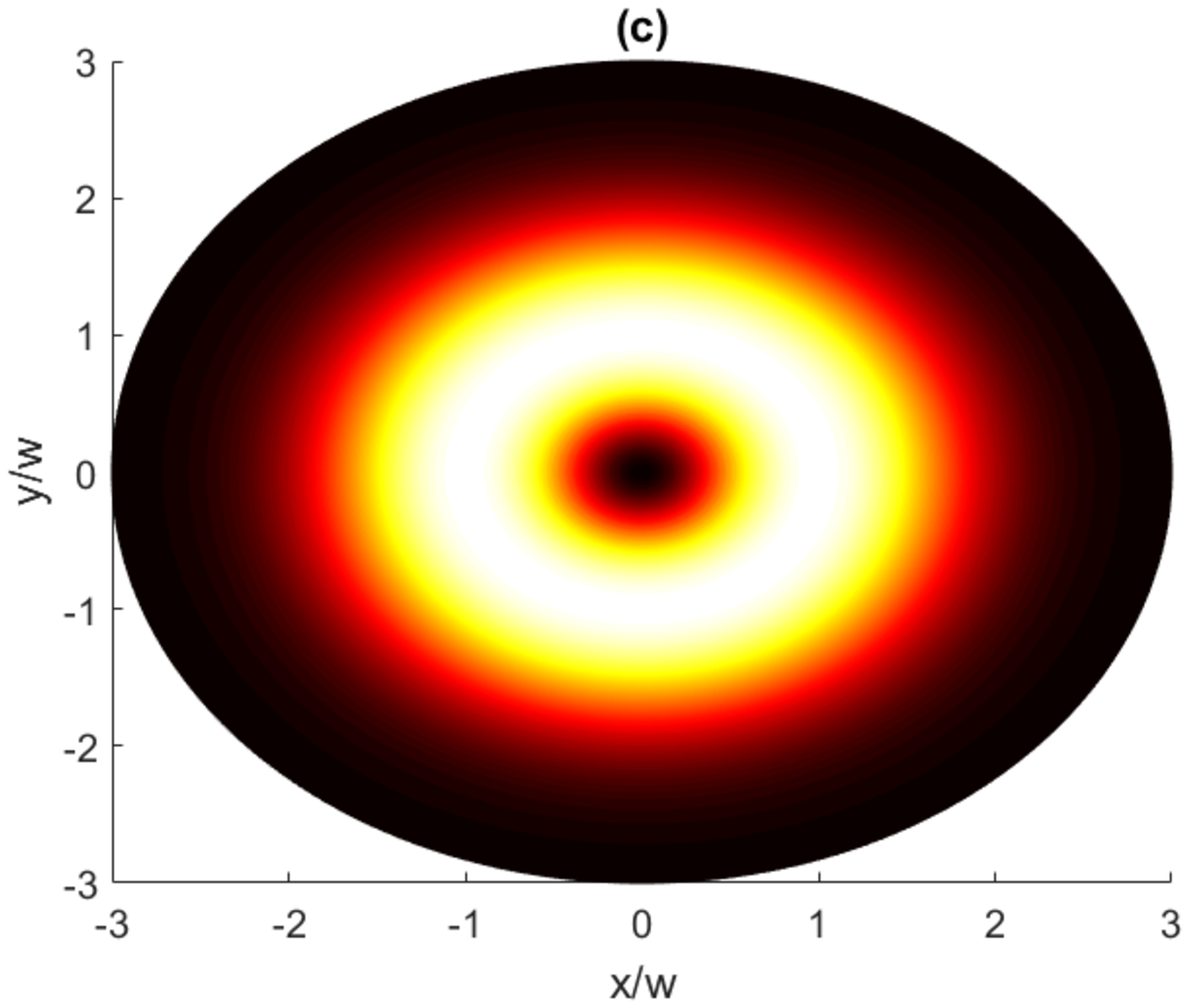} ~~\includegraphics[width=0.35\columnwidth]{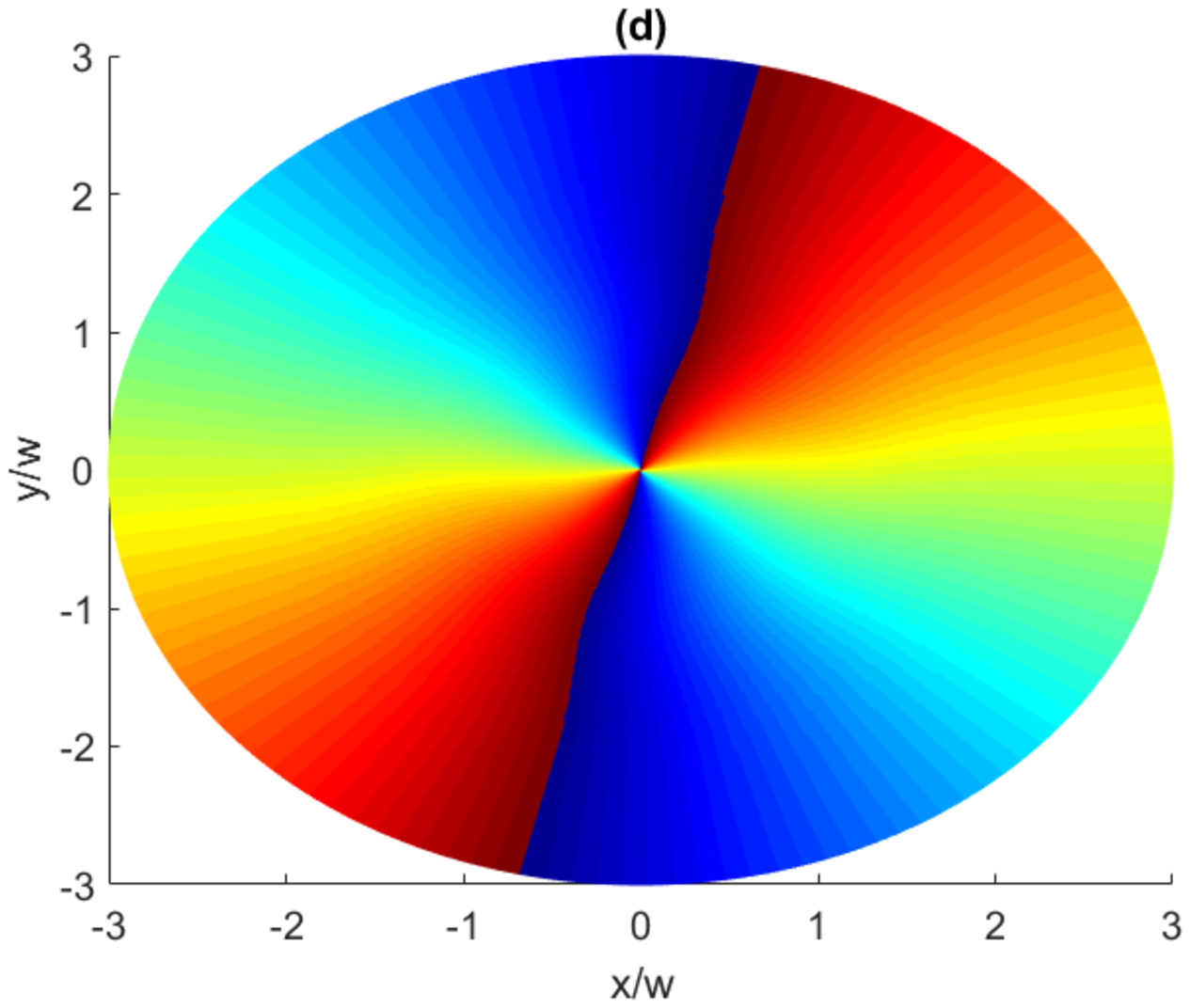} 

\includegraphics[width=0.35\columnwidth]{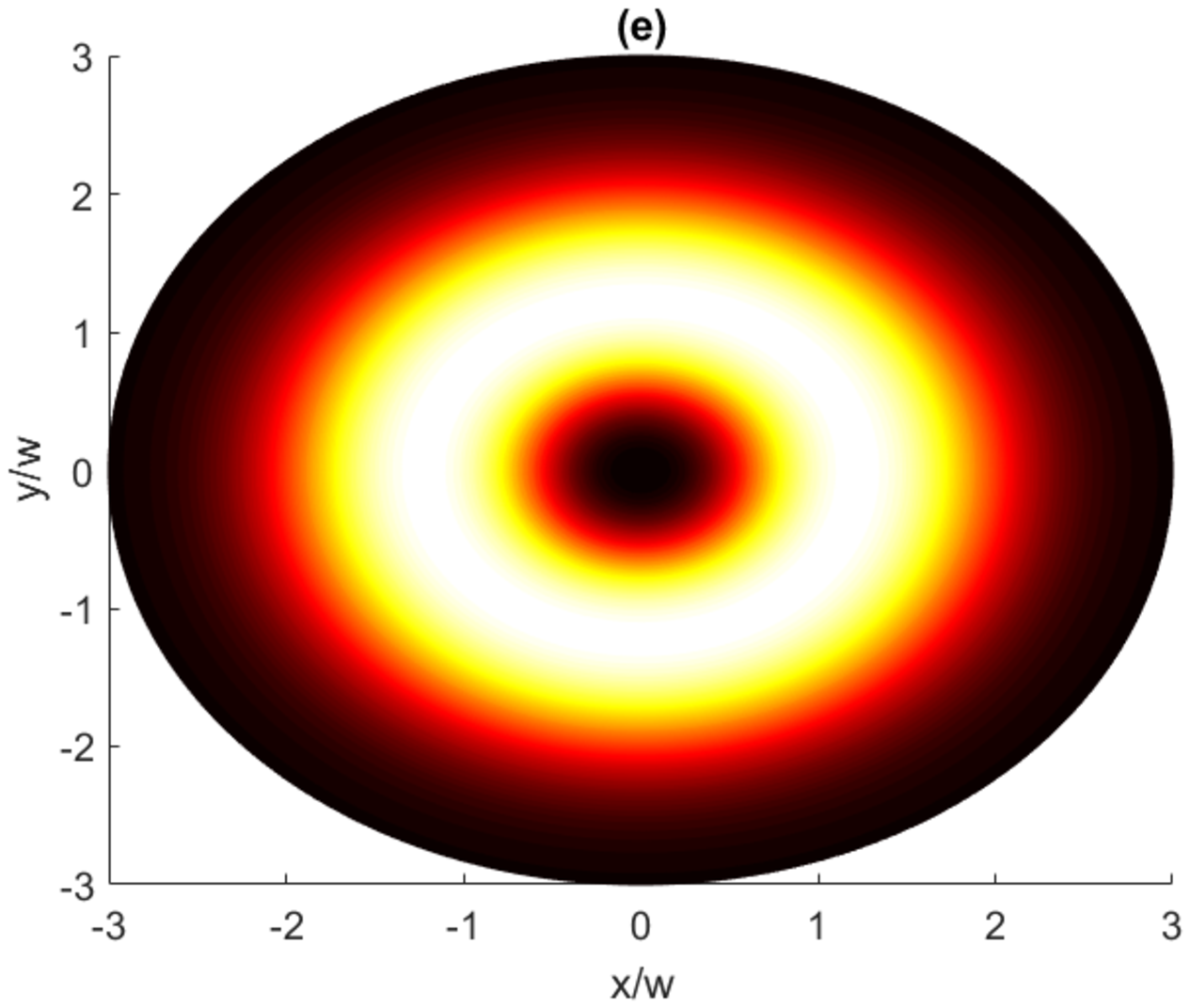} ~~\includegraphics[width=0.35\columnwidth]{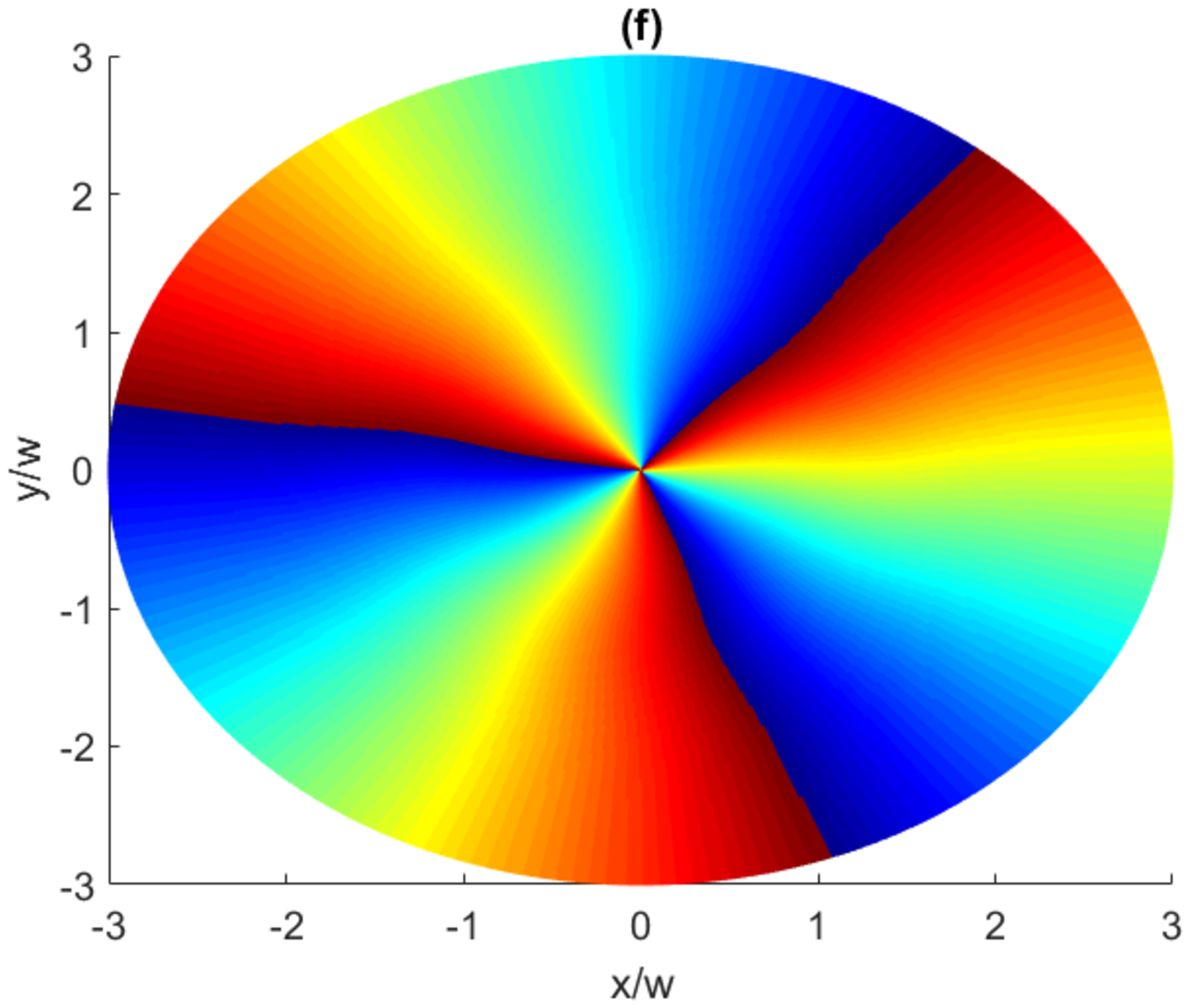} 

\caption{Intensity distributions (a, c, e) in arbitrary units as well as corresponding helical
phase patterns (b,d,f)) of the generated second probe vortex beam $\Omega_{p_{2}}$
with different OAM numbers $l=1$ (a,b), $l=2$ (c,d) and $l=3$ (e,f).
Here $\delta=11.9\Gamma$ (a, b), $\delta=11.53\Gamma$ (c, d), $\delta=11.83\Gamma$
(e, f) and the parameters are the same as Fig.~\ref{fig:3}. The position
is plotted in dimensionless units.}
\label{fig:5}
\end{figure}

The intensity distribution as well as the phase pattern profiles of
the generated vortex beam described by Eq.~\ref{eq:19} are plotted
in Fig.~\ref{fig:5} using Table~\ref{tab:Table1} and for different
OAM numbers $l=1,2$ and $3$. As can be seen in Fig.~\ref{fig:5}
(a, c, e), the intensity distributions are very similar to the patterns
displayed for the resonance one-photon detuning $\delta$ (Figs.~\ref{fig:3}(a,c,e)).
As illustrated in Fig.~~\ref{fig:5} (b, d, f), the phase patterns
are bended compared to the resonance case (Figs.~\ref{fig:3} (b,
d, f)). When $\delta$ is nonzero, the term $\exp[-i\alpha z/(2Ld)]$
in Eq.~\ref{eq:19} modulates the phase patterns since it contains
$|\Omega_{c_{2}}|$ which is not uniform in $(x,y)$ resulting in
bending of phase patterns. 

\section{Influence of phase mismatch}

In this Section we will investigate the influence of phase mismatch on the
exchange of optical vortices.  In order to include the effect of phase mismatch
we need to add an additional term to the Eq.~(\ref{eq:6}). We introduce the
geometrical phase mismatch 
\begin{equation}
\Delta_{k}=(k_{p_{1}}-k_{c_{1}}+k_{c_{2}}-k_{p_{2}})\cdot\hat{e}_{z},\label{eq:29}
\end{equation}
where $\hat{e}_{z}$ is the unit vector along the $z$ axis and $k_{p_{1}}$,
$k_{p_{2}}$, $k_{c_{1}}$ and $k_{c_{2}}$ are the wave-vectors of the light
beams. We consider the situation when all the light beams are co-propagating in
the same direction. Note, that the phase mismatch $\Delta_{k}$ can be minimized by
introducing a small angle between the propagation directions of the beams. For example, in Ref.
\cite{Lee2014} the value of $\Delta_{k}L=0.6$  has been achieved. In
the case of non-zero $\Delta_k$ instead of Eq.~(\ref{eq:6}) we have
\cite{Braje2004}
\begin{equation}
\frac{\partial\Omega_{p_{2}}}{\partial z}+c^{-1}\frac{\partial\Omega_{p_{2}}}{\partial t}+i\Delta_{k}\Omega_{p_{2}}=i\frac{\alpha\gamma_{e_{2}}}{2L}\rho_{e_{2}g}\,.\label{eq:30}
\end{equation}
Substituting
Eqs.~(\ref{eq:7}) and (\ref{eq:8}) into the Maxwell equations (\ref{eq:5})
and (\ref{eq:30}) and for $\Omega_{p_{2}}(0)=0$ we get
\begin{align}
\Omega_{p_{1}}(z)=&\frac{1}{s_{1}-s_{2}}\left((q_{1}-s_{2})\exp(s_{1}z)+(s_{1}-q_{1})\exp(s_{2}z)\right)\Omega_{p_{1}}(0),\label{eq:31}\\
\Omega_{p_{2}}(z)=&\frac{1}{q_{2}}\frac{s_{1}-q_{1}}{s_{1}-s_{2}}\left((q_{1}-s_{2})\exp(s_{1}z)+(s_{2}-q_{1})\exp(s_{2}z)\right)\Omega_{p_{1}}(0),\label{eq:32}
\end{align}
where
\begin{align}
q_{1}=&\frac{-i\alpha}{2Ld}\frac{|\Omega_{c_{2}}|^{2}}{|\Omega|^{2}},\label{eq:32-a}\\
q_{2}=&\frac{i\alpha}{2Ld}\frac{\Omega_{c_{1}}\Omega_{c_{2}}^{*}}{|\Omega|^{2}},\label{eq:32-b}
\end{align}
 and 
\begin{align}
s_{1}=&\frac{1}{2}\left(-(\frac{i\alpha}{2Ld}+i\Delta_{k})+\sqrt{(\frac{i\alpha}{2Ld}+i\Delta_{k})^{2}+4(\frac{\alpha\Delta_{k}}{2Ld}\frac{|\Omega_{c_{2}}|^{2}}{|\Omega|^{2}})}\right),\label{eq:33}\\
s_{2}=&\frac{1}{2}\left(-(\frac{i\alpha}{2Ld}+i\Delta_{k})-\sqrt{(\frac{i\alpha}{2Ld}+i\Delta_{k})^{2}+4(\frac{\alpha\Delta_{k}}{2Ld}\frac{|\Omega_{c_{2}}|^{2}}{|\Omega|^{2}})}\right).\label{eq:34}
\end{align}
One can show that when $\Delta_{k}=0$, Eqs.~(\ref{eq:31}) and (\ref{eq:32})
reduce to Eqs.~(\ref{eq:9}) and (\ref{eq:10}). However, the analytical
expressions given in Eqs.~(\ref{eq:31}) and (\ref{eq:32}) in presence
of the phase mismatch $\Delta_{k}$ are too complicated to see if
the OAM of the control field $\Omega_{c_{2}}$ is transferred to the
second generated probe beam $\Omega_{p_{2}}$. Hence, we follow the numerical approach to analyze whether or not the exchange of optical vortices is possible. 

When $z=L$ and using Eqs.~(\ref{eq:15})--(\ref{eq:17}) and
Eqs.~(\ref{eq:32-a})--(\ref{eq:32-b}) we can rewrite Eqs.~(\ref{eq:31}) and
(\ref{eq:32}) as
\begin{align}
\Omega_{p_{1}}(z=L)=&\frac{\Omega_{p_{1}}(0)}{S_{1}-S_{2}}\left[-\left(\frac{i\alpha}{d}\frac{\varepsilon_{c_{2}}^{2}\left(\frac{r}{w}\right)^{2|l|}\exp\left(-2\frac{r^{2}}{w^{2}}\right)}{|\Omega_{c_{1}}|^{2}+\varepsilon_{c_{2}}^{2}\left(\frac{r}{w}\right)^{2|l|}\exp\left(-2\frac{r^{2}}{w^{2}}\right)}+S_{2}\right)\exp(S_{1})\right.\nonumber\\
  &\left.+\left(\frac{i\alpha}{d}\frac{\varepsilon_{c_{2}}^{2}\left(\frac{r}{w}\right)^{2|l|}\exp\left(-2\frac{r^{2}}{w^{2}}\right)}{|\Omega_{c_{1}}|^{2}+\varepsilon_{c_{2}}^{2}\left(\frac{r}{w}\right)^{2|l|}\exp\left(-2\frac{r^{2}}{w^{2}}\right)}+S_{1}\right)\exp(S_{2})\right],\label{eq:37}\\
\Omega_{p_{2}}(z=L)=&\frac{-i\Omega_{p_{1}}(0)}{S_{1}-S_{2}}\frac{\left[S_{1}d\left(|\Omega_{c_{1}}|^{2}+\varepsilon_{c_{2}}^{2}\left(\frac{r}{w}\right)^{2|l|}\exp\left(-2\frac{r^{2}}{w^{2}}\right)\right)+i\alpha\varepsilon_{c_{2}}^{2}\left(\frac{r}{w}\right)^{2|l|}\exp\left(-2\frac{r^{2}}{w^{2}}\right)\right]}{\alpha|\Omega_{c_{1}}|\varepsilon_{c_{2}}\left(\frac{r}{w}\right)^{|l|}\exp\left(-\frac{r^{2}}{w^{2}}\right)\exp(-il\Phi)}\nonumber\\
&\times\left(S_{2}+\frac{i\alpha}{d}\frac{\varepsilon_{c_{2}}^{2}\left(\frac{r}{w}\right)^{2|l|}\exp\left(-2\frac{r^{2}}{w^{2}}\right)}{|\Omega_{c_{1}}|^{2}+\varepsilon_{c_{2}}^{2}\left(\frac{r}{w}\right)^{2|l|}\exp\left(-2\frac{r^{2}}{w^{2}}\right)}\right)(\exp(S_{2})-\exp(S_{1})),\label{eq:38}
\end{align}
where
\begin{align}
S_{1}=&-(\frac{i\alpha}{2d}+i\Delta_{k}L)+\sqrt{\left(\frac{i\alpha}{2d}+i\Delta_{k}L\right)^{2}+4\left(\frac{\alpha\Delta_{k}L}{2d}\frac{\varepsilon_{c_{2}}^{2}\left(\frac{r}{w}\right)^{2|l|}\exp\left(-2\frac{r^{2}}{w^{2}}\right)}{|\Omega_{c_{1}}|^{2}+\varepsilon_{c_{2}}^{2}\left(\frac{r}{w}\right)^{2|l|}\exp\left(-2\frac{r^{2}}{w^{2}}\right)}\right)},\label{eq:39}\\
S_{2}=&-(\frac{i\alpha}{2d}+i\Delta_{k}L)-\sqrt{\left(\frac{i\alpha}{2d}+i\Delta_{k}L\right)^{2}+4\left(\frac{\alpha\Delta_{k}L}{2d}\frac{\varepsilon_{c_{2}}^{2}\left(\frac{r}{w}\right)^{2|l|}\exp\left(-2\frac{r^{2}}{w^{2}}\right)}{|\Omega_{c_{1}}|^{2}+\varepsilon_{c_{2}}^{2}\left(\frac{r}{w}\right)^{2|l|}\exp\left(-2\frac{r^{2}}{w^{2}}\right)}\right)}.\label{eq:40}
\end{align}

\begin{figure}
\includegraphics[width=0.35\columnwidth]{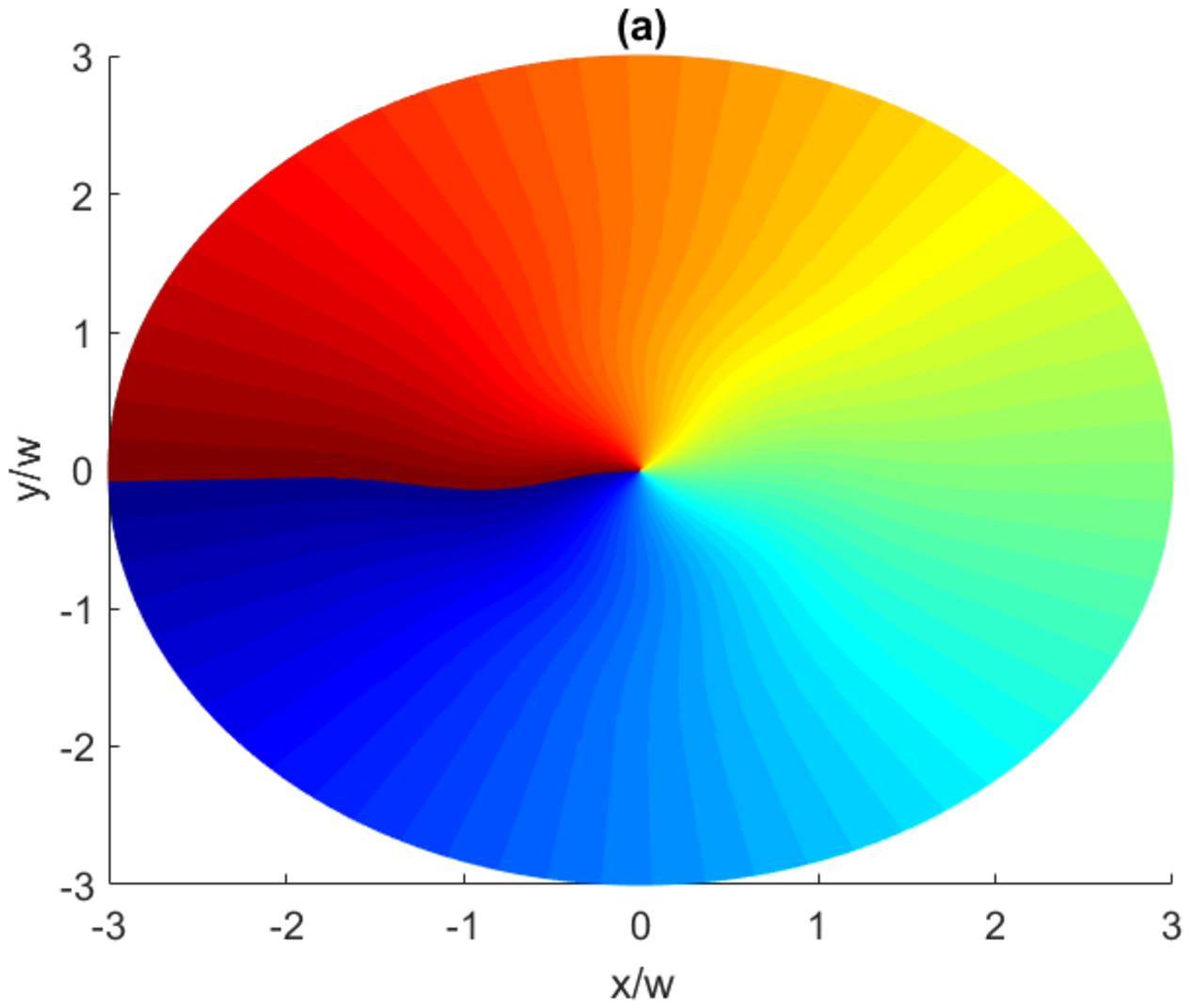} ~~\includegraphics[width=0.35\columnwidth]{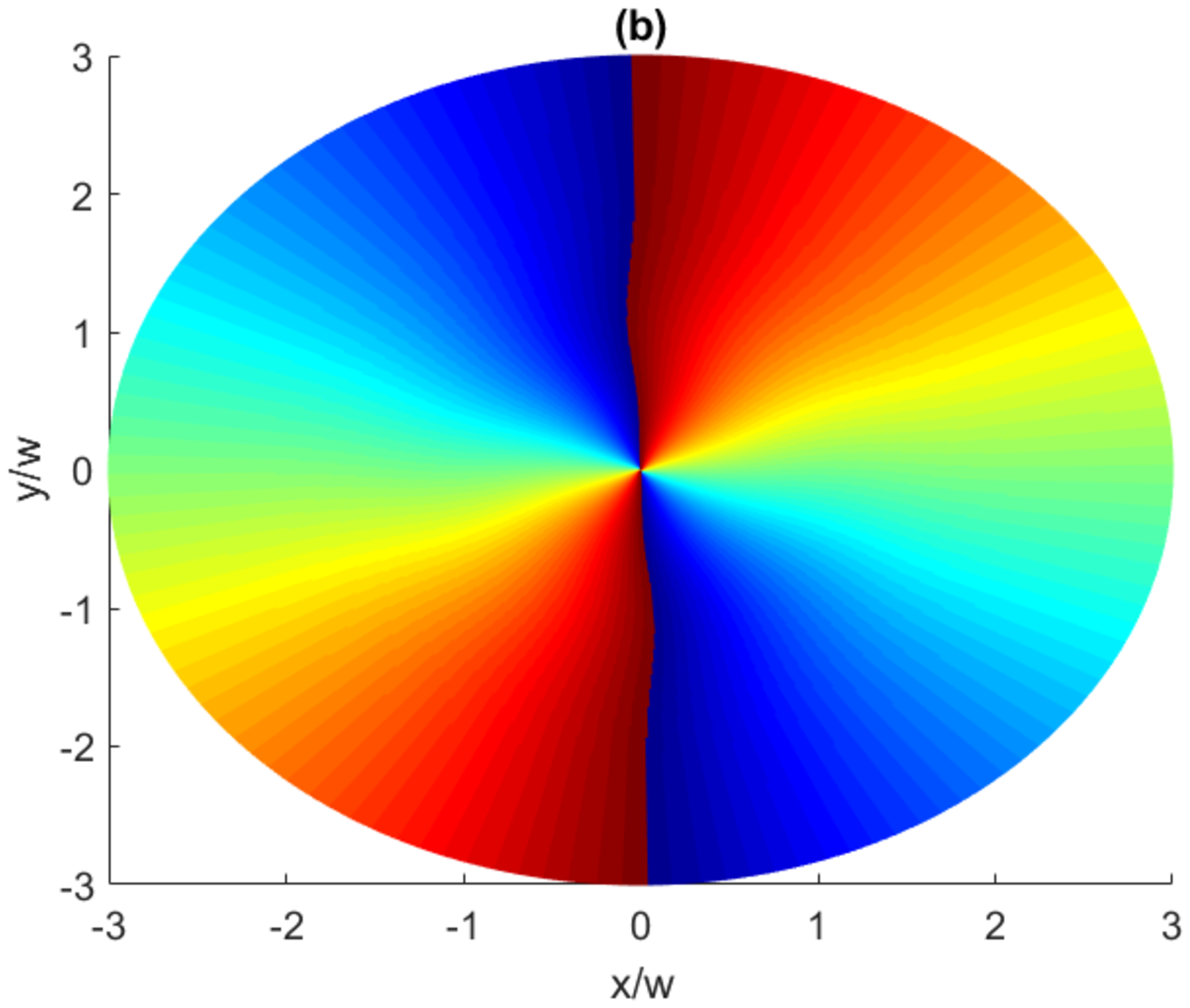} 

\includegraphics[width=0.35\columnwidth]{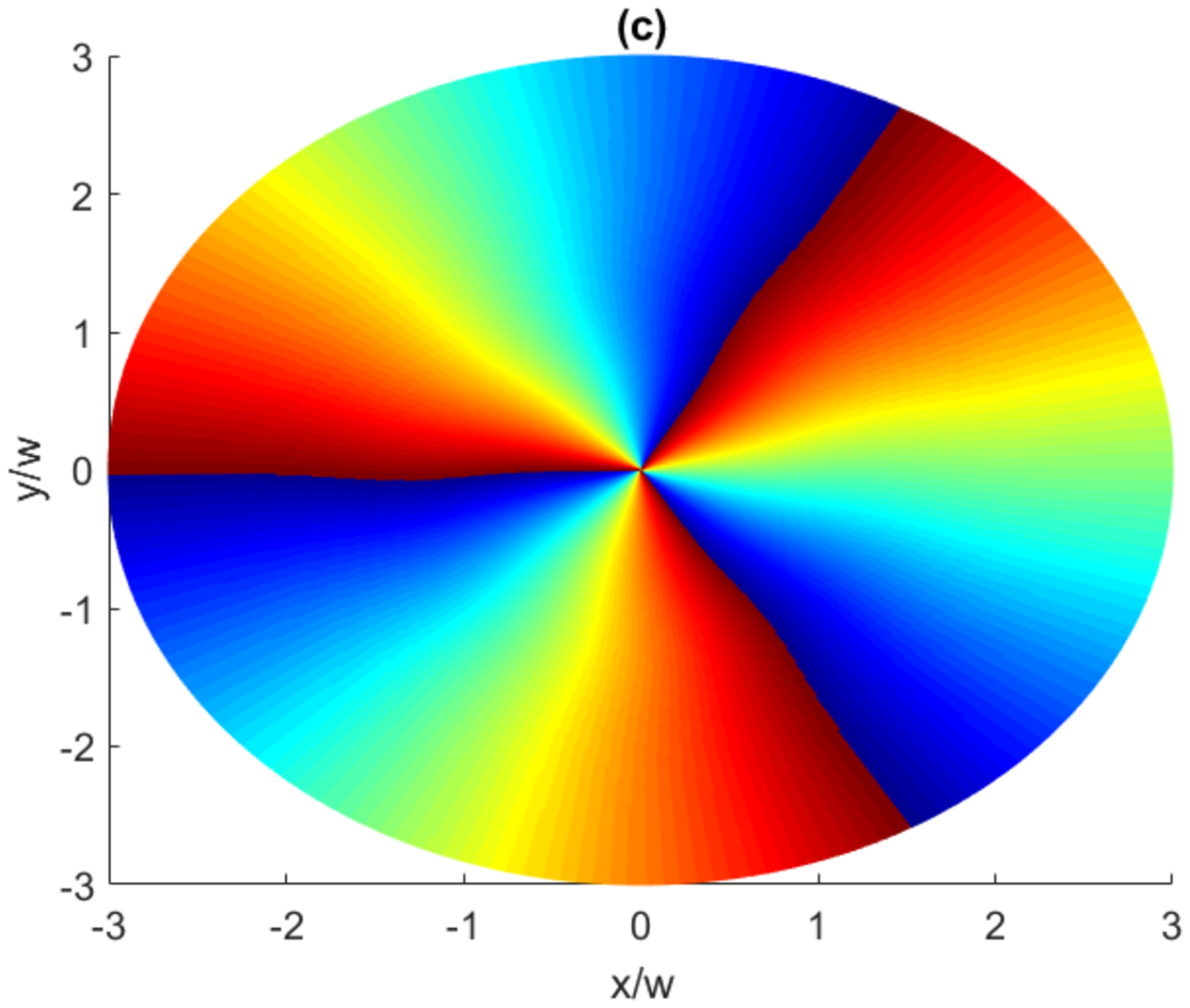} ~~\includegraphics[width=0.35\columnwidth]{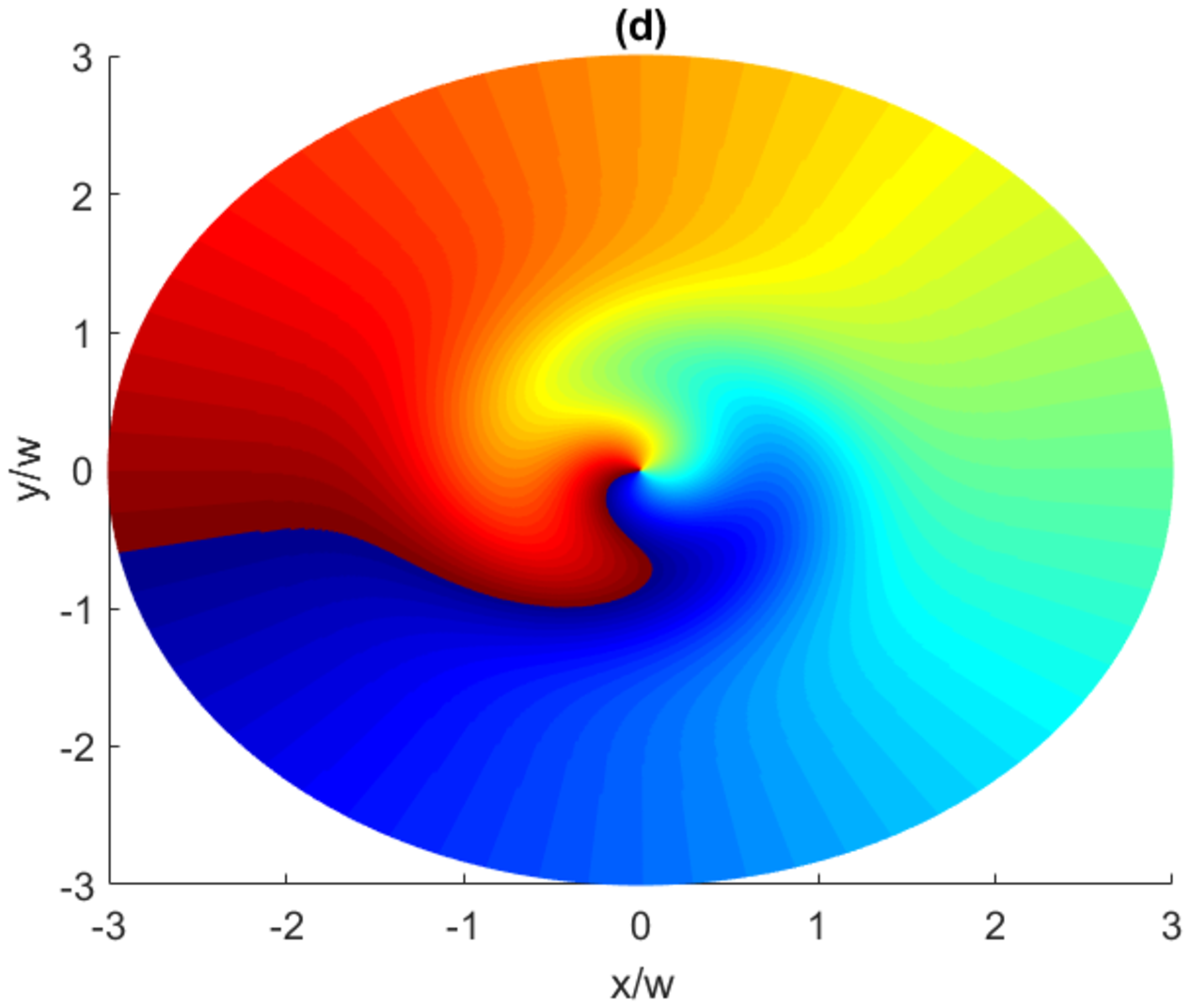} 

\includegraphics[width=0.35\columnwidth]{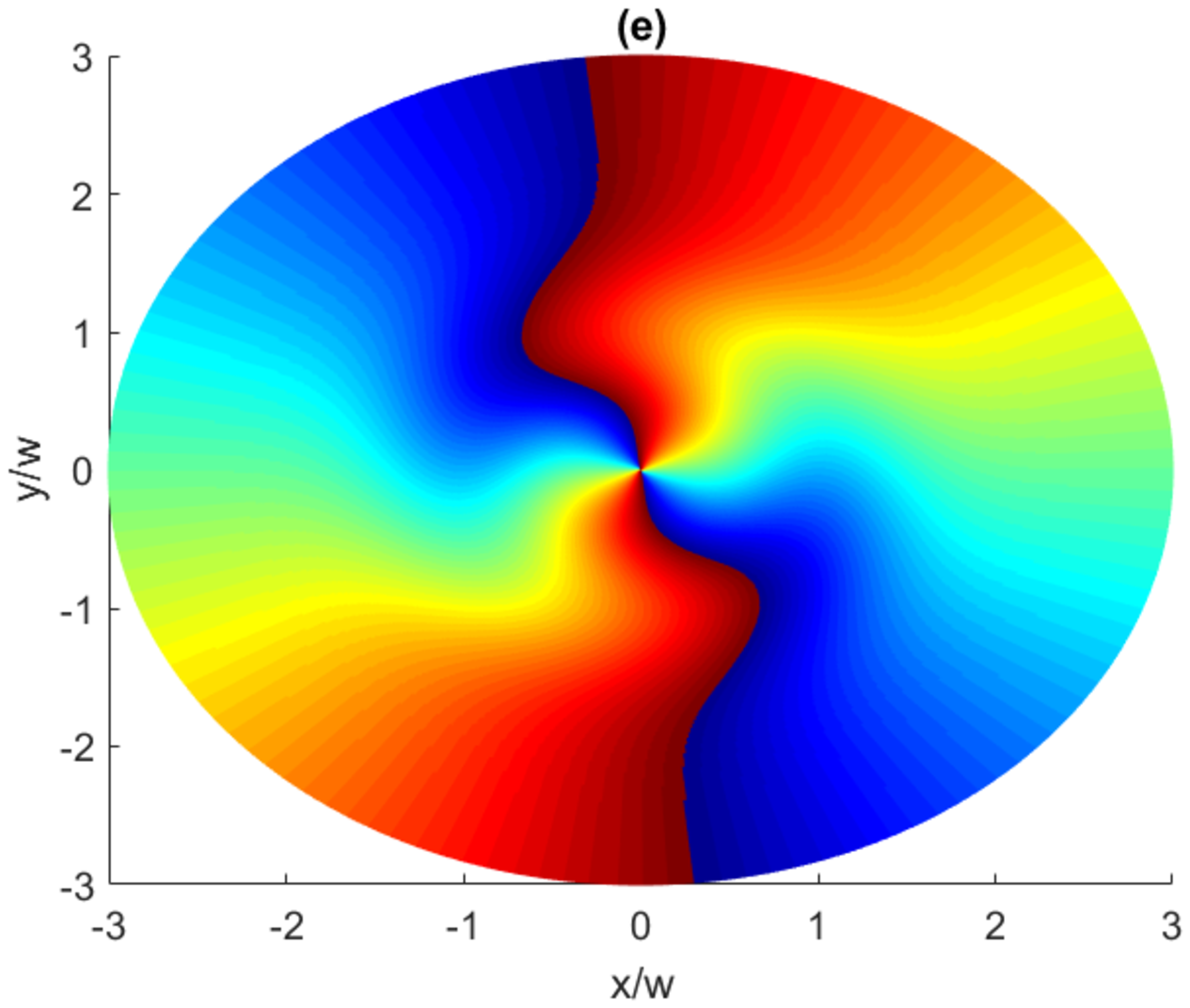} ~~\includegraphics[width=0.35\columnwidth]{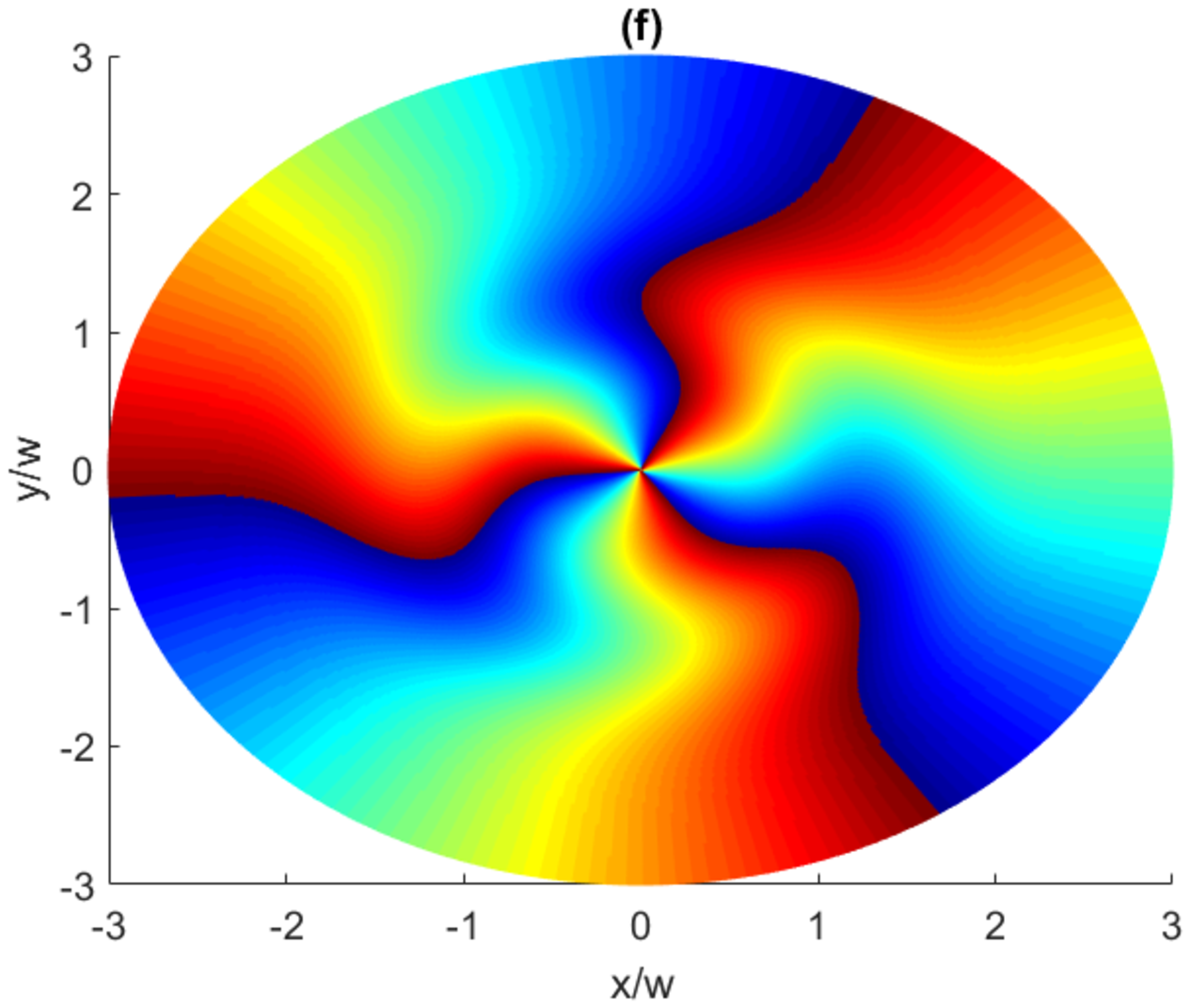} 

\caption{Helical phase patterns of the generated second probe vortex beam $\Omega_{p_{2}}$
described by Eq.~(\ref{eq:38}) for different OAM numbers $l=1$
(a,d), $l=2$ (b,e) and $l=3$ (c,f). Here $\Delta_{k}L=1$ (a, b,c)
and $\Delta_{k}L=10$ (d,e,f) and the parameters are the same as Fig.~\ref{fig:3}. The position
is plotted in dimensionless units.}
\label{fig:6}
\end{figure}

In Fig.~\ref{fig:6}, we plot the helical phase pattern of the generated
probe beam $\Omega_{p_{2}}(z=L)$ for $\Delta_{k}L=1$ (a, b,c) and
$\Delta_{k}L=10$ (d,e,f) and for different OAM numbers $l=1$ 
(a,d), $l=2$  (b,e) and $l=3$ 
(c,f). The singularity point appears obviously in phase patterns indicating
that the exchange of optical vortices is done and hence, the second
probe beam $\Omega_{p_{2}}$ has obtained the OAM of the control field
$\Omega_{c_{2}}$ in presence of the phase mismatch $\Delta_{k}$. 

\begin{figure}
\includegraphics[width=0.5\columnwidth]{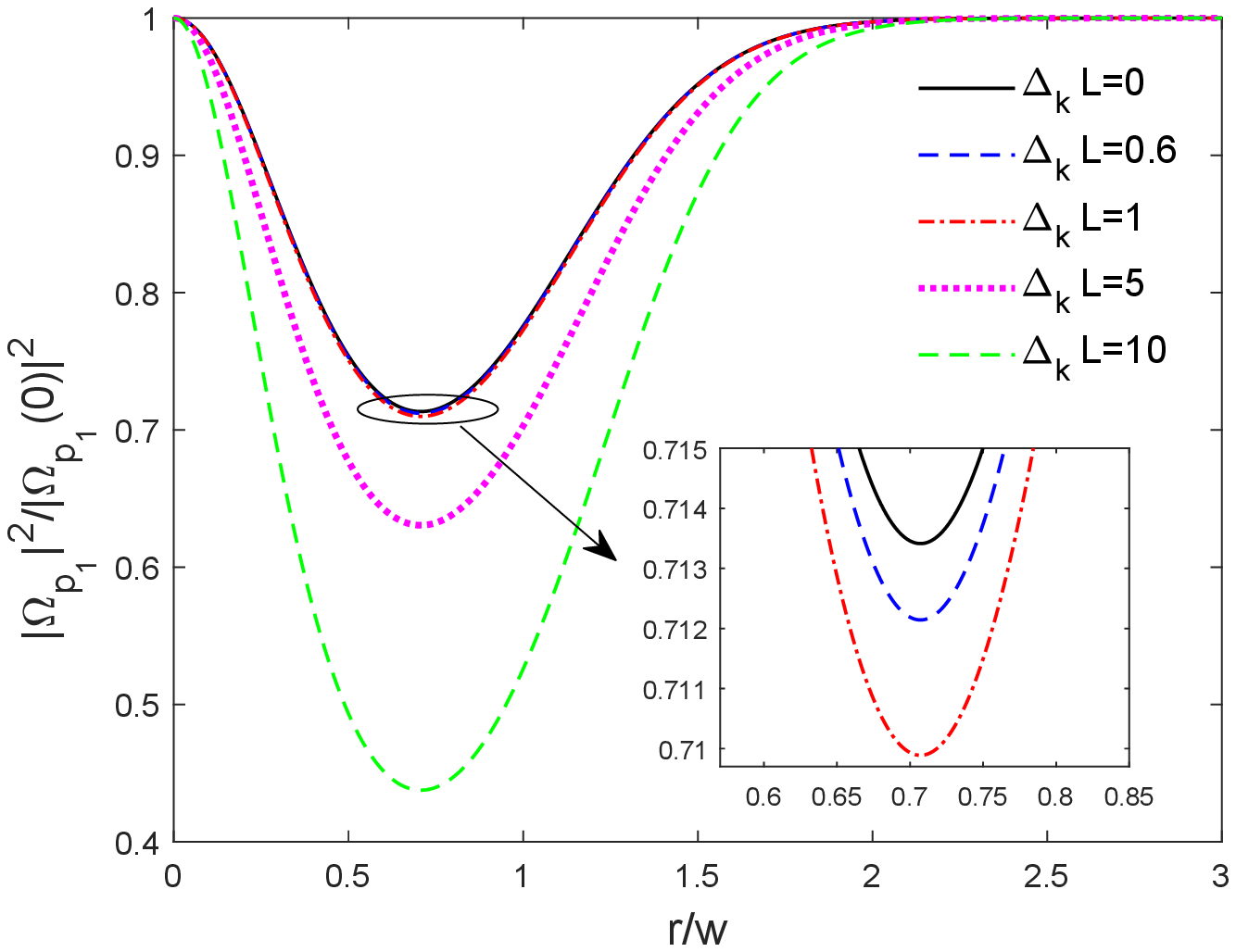} ~~\includegraphics[width=0.5\columnwidth]{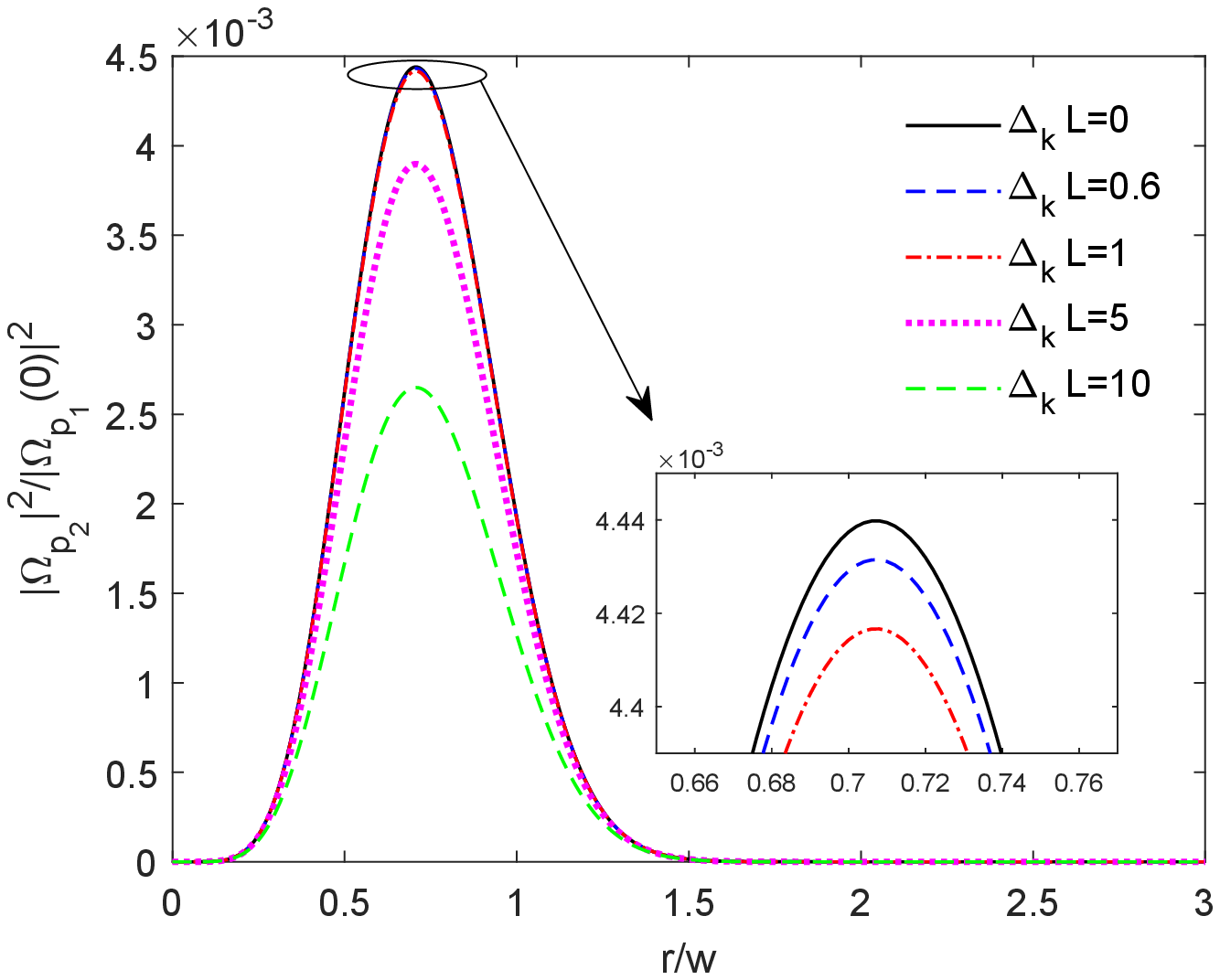} 

\caption{Dependence of the dimensionless quantities $|\Omega_{p_{1}}|^{2}/|\Omega_{p_{1}}(0)|^{2}$ (a)
and $|\Omega_{p_{2}}|^{2}/|\Omega_{p_{1}}(0)|^{2}$ (b) given in Eqs.~(\ref{eq:37})
and (\ref{eq:38}) on the dimensionless distance from the vortex core
$r/w$ for different values of $\Delta_{k}L$. Here, $l=1$ and the
parameters are the same as Fig.~\ref{fig:3}.}
\label{fig:7}
\end{figure}

In order to inspect the influence of the phase mismatch on efficiency
of OAM transfer, we plot in Fig.~\ref{fig:7} the intensities $|\Omega_{p_{1}}|^{2}/|\Omega_{p_{1}}(0)|^{2}$
and $|\Omega_{p_{2}}|^{2}/|\Omega_{p_{1}}(0)|^{2}$ in the whole range
of distance $r$ for different values of $\Delta_{k}L$. For small
values of $\Delta_{k}L$, the influence of phase mismatch is not significant,
yet they are seen to decrease the maximum amplitude of the second
generated probe beam for larger values of $\Delta_{k}L$. 

\section{Concluding Remarks\label{sec:concl}}

We have considered propagation of slow light with the OAM in a four-level
double-$\Lambda$ scheme of atom-light coupling. The medium is illuminated
by a pair of probe fields of weaker intensity as well as two control
fields with higher intensity. One of the control fields is allowed
to carry an OAM, while the second control field is a non vortex beam.
The intensity of one of the probe fields is zero at the entrance.
The generated probe field acquires the same OAM during the propagation.
Yet the presence of a non-vortex control beam makes the total intensity
of the control lasers not zero at the vortex core, preventing the
absorption losses. As a result, the OAM of the control field can be
transferred from the control field to a second generated probe field
through a FWM process and without switching on and off of the control
fields. Such a mechanism of OAM transfer is much simpler than the
previously considered double-tripod scheme where the exchange of vortices
is possible only when two control fields carry optical vortices of
opposite helicity. The energy losses during such an OAM transfer is
then calculated, and the analytical expression for the approximate
optimal one-photon detuning is obtained for which the efficiency of
generated probe field is maximum while the energy losses are minimal. 

Such an EIT based FWM setup can be implemented experimentally for
example using the $^{87}Rb$ atoms to form a DL level scheme. The
ground level $|g\rangle$ can then correspond to the $|5S_{1/2},F=1,m_{F}=0\rangle$
hyperfine state. The lower state $|s\rangle$ can be attributed to
the $|5S_{1/2},F=2,m_{F}=0\rangle$ state, whereas we can choose the
two excited states $|e_{1}\rangle$ and $|e_{2}\rangle$ as: $|e_{1}\rangle=|5P_{1/2},F=2,m_{F}=1\rangle$
and $|e_{2}\rangle=|5P_{3/2},F=2,m_{F}=1\rangle$.
\begin{acknowledgments}
This research was funded by the European Social Fund under grant No. 09.3.3-LMT-K-712-01-0051. H. R. H. gratefully acknowledges professor Lorenzo Marrucci and Filippo Cardano
for useful discussions and for providing advises on the OAM subject. 
\end{acknowledgments}

\end{document}